\documentclass[pra,aps,preprint,amsmath,amssymb,showpacs]{revtex4}
\usepackage{graphicx}
\usepackage{epsfig}
\usepackage{epstopdf}

\include{graphics}
\begin{document}


\title{Large stable oscillations due to Hopf bifurcations 
in amplitude dynamics of colliding soliton sequences}

\author{Avner Peleg$^{1}$, Debananda Chakraborty$^{2}$}

\affiliation{$^{1}$ Department of Exact Sciences, Afeka College of Engineering, 
Tel Aviv 69988, Israel 
\\
$^{2}$ Department of Mathematics, New Jersey City University, 
Jersey City, New Jersey 07305, USA}

\date{\today}

\begin{abstract}
We demonstrate that the amplitudes of optical solitons in nonlinear multisequence 
optical waveguide coupler systems with weak linear and cubic gain-loss  
exhibit large stable oscillations along ultra-long distances. 
The large stable oscillations are caused by supercritical Hopf bifurcations  
of the equilibrium states of the Lotka-Volterra (LV) models for dynamics of soliton amplitudes.   
The predictions of the LV models are confirmed by numerical simulations with 
the coupled cubic nonlinear Schr\"odinger (NLS) propagation models 
with $2 \le N \le 4$ pulse sequences. Thus, we provide the first demonstration of intermediate 
nonlinear amplitude dynamics in multisequence soliton systems, 
described by the cubic NLS equation. Our findings are also an important 
step towards realization of spatio-temporal chaos with multiple periodic sequences 
of colliding NLS solitons. 
\end{abstract}

\pacs{42.65.Tg, 42.65.Sf, 05.45.Yv}

\maketitle


\section{Introduction}
\label{Introduction}
The cubic nonlinear Schr\"odinger (NLS) equation, 
which describes wave propagation in the presence of 
second-order dispersion and cubic (Kerr) nonlinearity, 
is one of the most widely used nonlinear wave models 
in physics. It was successfully employed to describe 
water wave dynamics \cite{Zakharov84,Newell85}, 
Bose-Einstein condensates \cite{Dalfovo99,BEC2008}, 
and pulse propagation in nonlinear optical waveguides \cite{Agrawal2001,Mollenauer2006}.  
The fundamental NLS solitons are the most ubiquitous solutions of the cubic NLS equation 
due to their stability. Indeed, stable dynamics of single NLS solitons and of a 
single periodic sequence of NLS solitons has been observed in a wide range 
of physical systems \cite{Zakharov84,Newell85,Dalfovo99,BEC2008,Agrawal2001,Mollenauer2006}. 
However, the situation is very different for propagation of multiple periodic soliton sequences. 
Such multisequence propagation setups are of particular interest in nonlinear broadband (multichannel) 
optical waveguide systems \cite{MM98,Agrawal2001,Mollenauer2006}. In these waveguide systems, 
the solitons in each periodic sequence propagate with the same frequency and group velocity, 
but the frequency and group velocity are different 
for solitons from different sequences \cite{MM98,Agrawal2001,Mollenauer2006}. 
As a result, intersequence soliton collisions are frequent and can lead 
to significant amplitude and frequency shifts   
and to severe transmission degradation.
In fact, multichannel transmission with NLS solitons is unstable due to 
resonant emission of small-amplitude waves \cite{MM98,PNT2016,CPN2016}. 
Considering the ubiquity of the fundamental NLS soliton and of the single 
soliton sequence, the fact that stable long-distance propagation of multiple sequences 
of fundamental solitons has not been demonstrated in any system described 
by the cubic NLS equation is quite troubling. In particular, one would not expect 
intermediate or strongly nonlinear dynamics of soliton amplitudes 
in these systems.

In Refs. \cite{NP2010,PNC2010,PC2012,CPJ2013,NPT2015,PNT2016}, we developed a general method for 
stabilizing dynamics of soliton amplitudes in nonlinear multisequence optical waveguide systems 
with nonlinear dissipation. The method is based on showing that amplitude dynamics 
induced by nonlinear dissipation in 
$N$-sequence waveguide systems can be approximately  
described by $N$-dimensional Lotka-Volterra (LV) models. Stability analysis of the 
equilibrium states of the LV models can then be used for realizing 
stable amplitude dynamics along ultra-long distances.  
However, due to the inherent instability 
of mulitchannel soliton-based transmission against radiation emission, 
the distances along which stable amplitude dynamics was observed   
in numerical simulations were initially limited to a few 
hundred dispersion lengths \cite{PNC2010,PC2012}. 
Significant increase in the stable propagation distances was enabled by 
the introduction of frequency dependent linear gain-loss in 
$N$-waveguide couplers \cite{PNT2016,CPN2016}. 
The limiting cause for transmission instability in the latter systems was 
associated with radiation emission due to the effects of dissipative perturbations 
on single-soliton propagation \cite{PNT2016}. 
Therefore, this process is a serious obstacle for 
observing intermediate and strongly nonlinear amplitude dynamics 
in multichannel transmission with NLS solitons. Indeed, in all previous studies 
of multichannel soliton-based transmission, the dissipation-induced amplitude 
dynamics was only weakly nonlinear \cite{NP2010,PNC2010,PC2012,CPJ2013,NPT2015,PNT2016}. 
Furthermore, intermediate or strongly nonlinear amplitude dynamics 
has not yet been demonstrated in any multisequence soliton system, 
described by the cubic NLS equation.

A common mechanism for inducing intermediate nonlinear dynamics 
is by means of supercritical Hopf bifurcations 
\cite{Holmes83,Lakshmanan2002,Field85,Murray89}. 
In this case, as the value of a physical parameter is changed beyond some 
threshold value, a stable equilibrium state of the dynamical model becomes unstable, 
and a stable limit cycle about the unstable equilibrium state appears \cite{Holmes83,Lakshmanan2002}.  
As a result, for parameter values 
larger than the threshold value, the system exhibits stable oscillations 
with relatively large amplitudes, i.e., intermediate nonlinear amplitude dynamics. 
Supercritical Hopf bifurcations occur in many physical systems, 
including electric circuits \cite{Lakshmanan2002}, 
chemical reactions \cite{Field85,Murray89,Wyman89},
and population dynamics \cite{Murray89,Odell1980,Arneodo80,Arneodo82,Namba2005,Vano2006,Previte2013}.
Here, we are interested in LV models, 
which describe dynamics of population sizes \cite{Murray89,Lotka25,Volterra28}
as well as the time evolution of chemical concentrations in 
certain chemical reactions \cite{Field85,Murray89,Wyman88,Wyman89,Li2008}. 
The occurrence of supercritical Hopf bifurcations 
in LV models is of special interest, since in some cases, as the value 
of the bifurcation parameter is further changed, the limit cycle undergoes 
a period doubling cascade, and finally, chaotic dynamics is observed
\cite{Arneodo80,Arneodo82,Wyman89,Namba2005,Vano2006}.

In the current paper, we provide the first demonstration of intermediate 
nonlinear dynamics of soliton amplitudes in multisequence soliton systems, 
described by the cubic NLS equation. For this purpose, we study propagation 
of multiple periodic soliton sequences in nonlinear optical waveguide coupler   
systems with weak linear gain-loss, weak broadband cubic gain-loss, and narrowband 
Kerr nonlinearity. The values of the gain-loss coefficients are chosen such 
that the equilibrium states of the LV models for amplitude dynamics 
undergo supercritical Hopf bifurcations. This enables observation of    
large stable oscillations of soliton amplitudes along ultra-long distances.  
The narrowband nature of the Kerr nonlinearity and the broadband nature 
of the cubic gain-loss lead to enhanced pulse pattern stability compared 
with the waveguides considered in 
Refs. \cite{NP2010,PNC2010,PC2012,CPJ2013,NPT2015,PNT2016}.
Since two of the LV models that we study exhibit chaotic dynamics, 
our findings are an important step towards realization of spatio-temporal chaos 
with multiple sequences of colliding NLS solitons.

The rest of the paper is organized as follows. 
In Section \ref{models}, we present the coupled-NLS models for 
pulse propagation and the LV models for dynamics of soliton amplitudes. 
In Section \ref{setups}, we present four examples for multisequence 
waveguide coupler systems, in which the soliton amplitudes exhibit 
large stable oscillations along ultra-long distances. For each of the four systems 
we present the predictions of the LV models, the results of numerical 
simulations with the coupled-NLS models, and a comparison.   
Our conclusions are presented in Section \ref{conclusions}.

\section{Coupled-NLS and LV models}
\label{models}

\subsection{Coupled-NLS propagation models}
\label{NLS_models}
We consider propagation of $N$ sequences of optical pulses in an optical waveguide 
coupler, consisting of $N$ close waveguides, where each sequence propagates 
through its own waveguide. We assume a multisequence setup, where the pulses in each 
sequence propagate with the same group velocity, but where 
the group velocity is different for pulses from different sequences \cite{Agrawal2001,MM98,Mollenauer2006}.  
Additionally, we assume that the sequences propagate in the presence of 
second-order dispersion, Kerr nonlinearity, and weak linear and cubic gain-loss.
Thus, the propagation is described by the following system 
of $N$ coupled cubic NLS equations \cite{Agrawal2001,Agrawal2007a,PNC2010,CPN2016}: 
\begin{eqnarray} &&
i\partial_z\psi_{j}+\partial_{t}^2\psi_{j}+2|\psi_{j}|^2\psi_{j}=
i{\cal F}^{-1}(G_j(\omega,z) \hat\psi_{j})/2
\nonumber \\&&
-2i\sum_{k=1}^{N}(1-\delta_{jk})\epsilon_{3jk}
|\psi_{k}|^2\psi_{j},  
\label{HB1}
\end{eqnarray}     
where $\psi_{j}$ is the envelope of the electric field of the $j$th sequence, 
$1 \le j \le N$, $z$ is propagation distance, $t$ is time, and $\omega$ is frequency \cite{dimensions}. 
In Eq. (\ref{HB1}), $G_j(\omega,z)$ is the linear gain-loss 
experienced by $j$th sequence pulses, 
$\hat\psi_{j}$ is the Fourier transform of $\psi_{j}$ with respect to time, 
${\cal F}^{-1}$ is the inverse Fourier transform, 
and $\delta_{jk}$ is the Kronecker delta function.
The coefficients $\epsilon_{3jk}$, which describe the strength of cubic gain-loss interaction 
between $j$th and $k$th sequence pulses, satisfy $|\epsilon_{3jk}| \ll 1$. 
The second and third terms on the left hand side 
of Eq. (\ref{HB1}) describe second-order dispersion effects  
and intrasequence interaction due to Kerr nonlinearity.  
The first term on the right hand side of Eq. (\ref{HB1}) describes  
the effects of frequency dependent linear gain-loss,  
while the second term corresponds to intersequence 
interaction due to cubic gain-loss.  
We assume that Kerr nonlinearity is narrowband, i.e., that it is 
negligible for frequency differences that are much larger than the spectral 
width of the pulses. In addition, we assume that cubic gain-loss is broadband, 
i.e., that it is non-negligible only for frequency differences that are much larger than the 
spectral width of the pulses. As a result, we can neglect interchannel 
interaction due to Kerr nonlinearity and intrachannel interaction due to cubic gain-loss. 
These properties, which are new compared with the waveguides   
studied in all previous works on multichannel soliton-based transmission 
\cite{PNC2010,PC2012,PNT2016,CPN2016,CPJ2013,NPT2015},
lead to significant enhancement of pulse pattern stability 
and enable the observation of large stable oscillations of pulse amplitudes 
along ultra-long distances in simulations with Eq. (\ref{HB1}).

The $k$th pulse in the $j$th sequence 
is a fundamental soliton of the unperturbed NLS equation 
$i\partial_z\psi_{jk}+\partial_{t}^2\psi_{jk}+2|\psi_{jk}|^2\psi_{jk}=0$. 
The envelope of this soliton is  
$\psi_{sjk}(t,z)=\eta_{j}\exp(i\chi_{jk})\mbox{sech}(x_{jk})$,
where $x_{jk}=\eta_{j}\left(t-y_{jk}-2\beta_{j} z\right)$, 
$\chi_{jk}=\alpha_{j}+\beta_{j}(t-y_{jk})+
\left(\eta_{j}^2-\beta_{j}^{2}\right)z$, 
and $\eta_{j}$, $\beta_{j}$, $y_{jk}$, and $\alpha_{j}$ 
are the soliton amplitude, frequency, position, and phase.      
In a periodic sequence with index $j$, the positions of the 
$k$th and $(k+1)$th pulses in the sequence are related by 
$y_{jk}=y_{jk-1}+T$, where $T$ is the intrasequence separation 
between adjacent pulses.

The form of the linear gain-loss $G_j(\omega,z)$ is chosen such that 
large stable oscillations of soliton amplitudes are enabled, while pulse pattern 
destabilization due to radiation emission is suppressed. 
In particular, we choose a form 
similar to the one that was used in Refs. \cite{CPN2016,PNT2016,PNT2017}: 
\begin{eqnarray} &&
G_j(\omega,z) \!=\!
\left\{ \begin{array}{l l}
\epsilon_{1} g_{j}(z) &  \mbox{ if $\beta_{j}(0)-W/2 < \omega \le \beta_{j}(0)+W/2$,}\\
-g_{L} &  \mbox{elsewhere,}\\
\end{array} \right. 
\label{HB2}
\end{eqnarray}     
where $\epsilon_{1}$ is the linear gain coefficient, $0<\epsilon_{1} \ll 1$,  
$\beta_{j}(0)$ is the initial frequency of 
$j$th sequence solitons, and $g_{L}$ is an $O(1)$ positive constant. 
The spectral width $W$ in Eq. (\ref{HB2}) satisfies $1 < W \le \Delta\beta$, 
where the frequency spacing $\Delta\beta$ is defined by:  
$\Delta\beta=\beta_{j+1}(0)-\beta_{j}(0)$ for $1 \le j \le N-1$.
The function $g_{j}(z)$ is: $g_{j}(z)=g_{1j}+g_{2j}\eta_{j}(z)+g_{3j}\eta_{j}^{2}(z)$, 
where $\eta_{j}(z)$ is the amplitude of $j$th sequence solitons.   
The values of the constants $g_{1j}$, $g_{2j}$, and $g_{3j}$ are chosen such that 
the soliton amplitudes exhibit large stable oscillations due to Hopf bifurcations of 
the equilibrium states of the LV models for amplitude dynamics. 
The strong linear loss $g_{L}$ leads to efficient suppression of 
instability due to emission of radiation with frequencies 
outside the interval $(\beta_{j}(0)-W/2, \beta_{j}(0)+W/2]$. 
Simulations with Eq. (\ref{HB1}) show that efficient 
mitigation of radiative instability is achieved for $g_{L}$ and $W$ values 
around 0.5 and 10, respectively.     
The flat gain in the interval $(\beta_{j}(0)-W/2, \beta_{j}(0)+W/2]$ 
can be realized by flat-gain amplifiers \cite{Becker99}, 
and the strong loss outside of this interval can be achieved by filters \cite{Becker99} 
or by waveguide impurities \cite{Agrawal2001}.

\subsection{LV models for dynamics of soliton amplitudes}
\label{LV_models}
In Refs. \cite{NP2010,PNC2010,PC2012,CPJ2013,NPT2015,PNT2016}, 
we showed that amplitude dynamics of $N$ periodic sequences of colliding solitons  
in nonlinear optical waveguides with weak dissipation can be 
described by $N$-dimensional LV models. 
The derivation of the LV models was based on the following assumptions. 
(1) The intrasequence separation $T$ satisfies: $T \gg 1$. 
In addition, the amplitudes are equal for all solitons from the 
same sequence, but are not necessarily equal for solitons from different sequences. 
(2) The sequences are either (a) subject to periodic 
temporal boundary conditions or (b) infinitely long. 
Setup (a) corresponds to waveguide-loop experiments 
and setup (b)  approximates long-haul transmission.
(3) As $T\gg 1$, intrasequence interaction is exponentially weak and is neglected. 
(4) High-order radiation emission effects are also neglected.

Under these assumptions, the soliton sequences remain periodic and therefore   
the amplitudes of all pulses in a given sequence 
follow the same dynamics. Taking into account collision-induced amplitude 
shifts due to cubic gain-loss and single-pulse amplitude changes 
due to linear gain-loss, we obtain the following equation for amplitude dynamics 
of $j$th sequence solitons \cite{PNC2010}: 
\begin{eqnarray} &&  
\frac{d\eta _{j}}{dz}=
\eta_{j}\left[\epsilon_{1}g_{j}(z)
-\frac{8}{T}\sum_{k=1}^{N}(1-\delta_{jk})\epsilon_{3jk}\eta_{k}\right]. 
\label{HB3}
\end{eqnarray} 
Equation (\ref{HB3}) has the form of a LV model for $N$ species \cite{Lotka25,Volterra28}.                                                                                                                                  
The choice of physical parameter values in Eq. (\ref{HB3}) is guided by the 
following requirements. (a) The LV model (\ref{HB3}) has an equilibrium state 
with equal or near-equal amplitudes, whose values are close to 1, when the value
of the bifurcation parameter $\mu$ is close to its Hopf bifurcation value $\mu_{H}$ \cite{mu}. 
(b) The equilibrium state with near-equal amplitudes undergoes a supercritical Hopf bifurcation 
at  $\mu=\mu_{H}$. As a result, the equilibrium state 
changes from a stable focus to an unstable state and a stable limit 
cycle around the unstable state appears. The emergence of the stable limit cycle is 
the key factor in enabling large stable amplitude oscillations. 
Requirement (a) is added so that the perturbation procedure 
leading to Eq. (\ref{HB3}) is valid when $\mu$ is close to $\mu_{H}$.       
Requirements (a) and (b) are new features of the LV model 
and the corresponding waveguide systems and were not considered 
in Refs. \cite{NP2010,PNC2010,PC2012,CPJ2013,NPT2015,PNT2016}.


\section{Specific multisequence transmission setups leading to intermediate 
nonlinear amplitude dynamics}
\label{setups}     

\subsection{Introduction}
\label{setups_intro}
We consider four multisequence waveguide coupler 
systems with weak linear and cubic gain-loss   
as prototypical examples for waveguide systems,  
where dynamics of soliton amplitudes with large stable oscillations 
that is induced by a supercritical Hopf bifurcation can be observed.  
For each waveguide coupler system, we present the predictions of the LV model (\ref{HB3}), 
the results of numerical simulations with the coupled-NLS model (\ref{HB1}), 
and a comparison.

Equation (\ref{HB1}) is numerically 
solved using the split-step method with periodic boundary conditions \cite{Agrawal2001,Yang2010}. 
Due to the periodic boundary conditions, the simulations describe propagation in a closed waveguide loop. 
The initial condition consists of $N$ periodic sequences 
of $2K+1$ solitons with amplitudes $\eta_{j}(0)$, 
frequencies $\beta_{j}(0)$, and zero phases:  
\begin{eqnarray} &&
\psi_{j}(t,0)\!=\!\sum_{k=-K}^{K}
\frac{\eta_{j}(0)\exp[i\beta_{j}(0)(t-kT)]}
{\cosh[\eta_{j}(0)(t-kT)]},
\label{HB4}
\end{eqnarray}
where $1\le j \le N$, and $2 \le N \le 4$. We use $W=10$ and $g_{L}=0.5$ 
for the parameters of the linear gain-loss function $G_j(\omega,z)$. 
For concreteness, we present here the results of simulations with 
$T=20$, $\Delta\beta=40$, $K=1$, and a final distance $z_{f}=5000$. 
We emphasize, however, that similar results are obtained with other physical parameter values.

\subsection{Two-sequence transmission}
\label{2D}  
The two-dimensional (2D) LV model for two-sequence transmission is a variant of the 
predator-prey model analyzed by Odell in Ref. \cite{Odell1980} in the context of population dynamics.  
The model is given by Eq. (\ref{HB3}) with  
$N\!=\!2$, $g_{11}\!=\!-\mu$, $g_{21}\!=\!g_{31}\!=\!0$, $g_{12}\!=\!-1$, 
$g_{22}\!=\!4$, $g_{32}\!=\!-2$, $\epsilon_{312}\!=\!-\epsilon_{1}T/8$,      
$\epsilon_{321}\!=\!\epsilon_{1}T/8$, where $\mu$ is a bifurcation parameter. 
Therefore, the LV model is:                 
\begin{eqnarray} &&  
\frac{d\eta _{1}}{dz}=
\epsilon_{1}\eta_{1}\left(-\mu+\eta_{2}\right), 
\nonumber \\&&
\frac{d\eta _{2}}{dz}=
\epsilon_{1}\eta_{2}\left(-1+4\eta_{2}-2\eta_{2}^{2}-\eta_{1}\right). 
\label{HB5}
\end{eqnarray}                                                                                                                 
Note that in this transmission system, the $j=1$ sequence propagates in the presence 
of linear loss with coefficient $g_{1}\!=\!-\mu$ and cubic gain 
intersequence interaction, while the $j=2$ sequence propagates 
in the presence of linear gain-loss with coefficient 
$g_{2}(z)\!=\!-1+4\eta_{j}-2\eta_{j}^{2}$ and cubic loss intersequence interaction. 
The equilibrium state with near-equal amplitudes 
is $(\eta_{1}^{(eq)},\eta_{2}^{(eq)})\!=\!(-1+4\mu-2\mu^2 ,\mu)$. This state undergoes 
a supercritical Hopf bifurcation as  $\mu$ changes through 1. 
For $\mu<1$, the equilibrium state is an unstable focus, 
which is surrounded by a stable limit cycle, while for $\mu>1$, it is a stable 
focus and the limit cycle does not exist \cite{Odell1980}.           
Thus, we expect large stable amplitude oscillations 
for $\mu<1$, and stable oscillations that decay to 
$\eta_{j}^{(eq)}$ for $\mu>1$.

\begin{figure}[htbp]
\centerline{\includegraphics[width=.75\columnwidth]{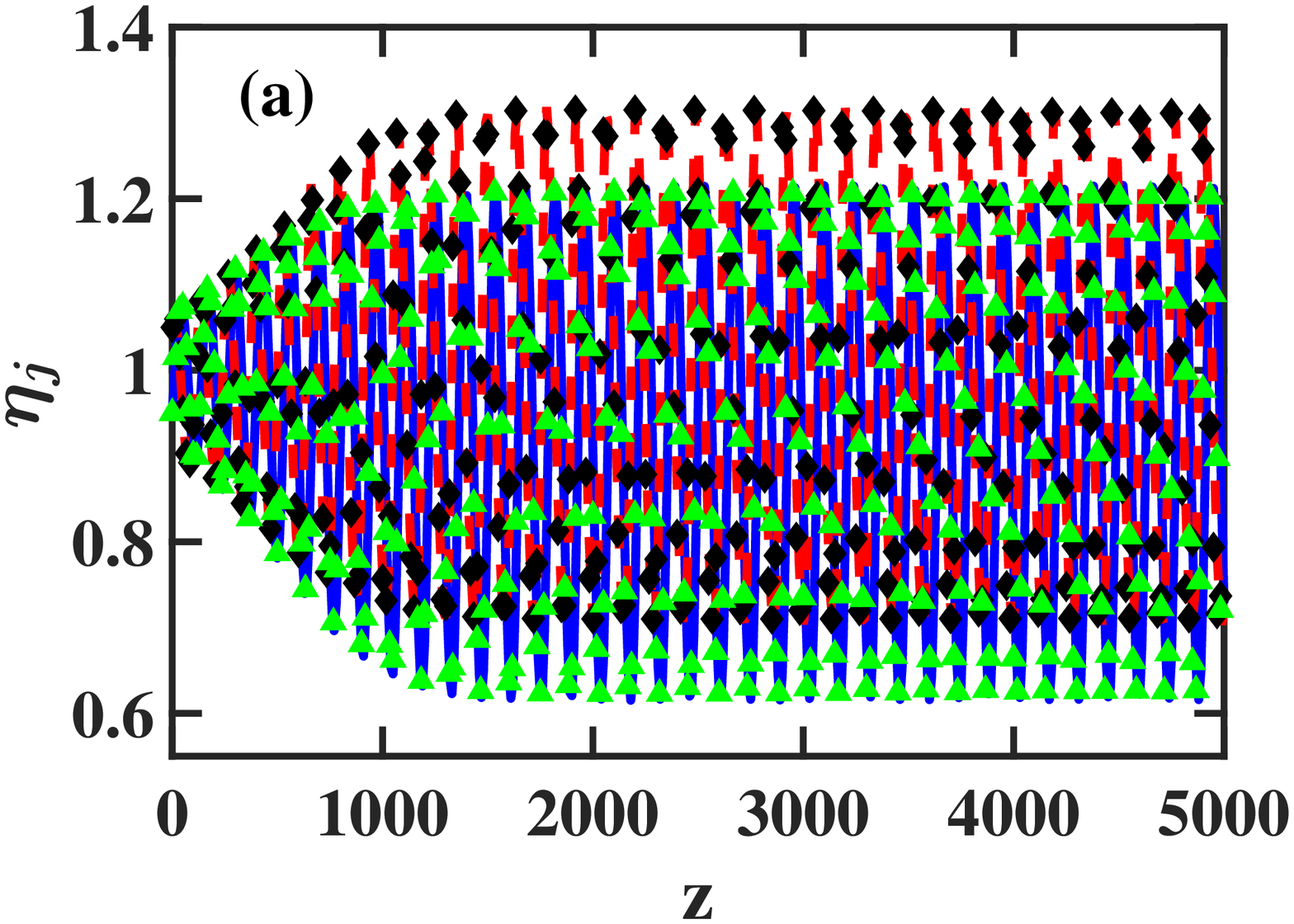}}
\centerline{\includegraphics[width=.75\columnwidth]{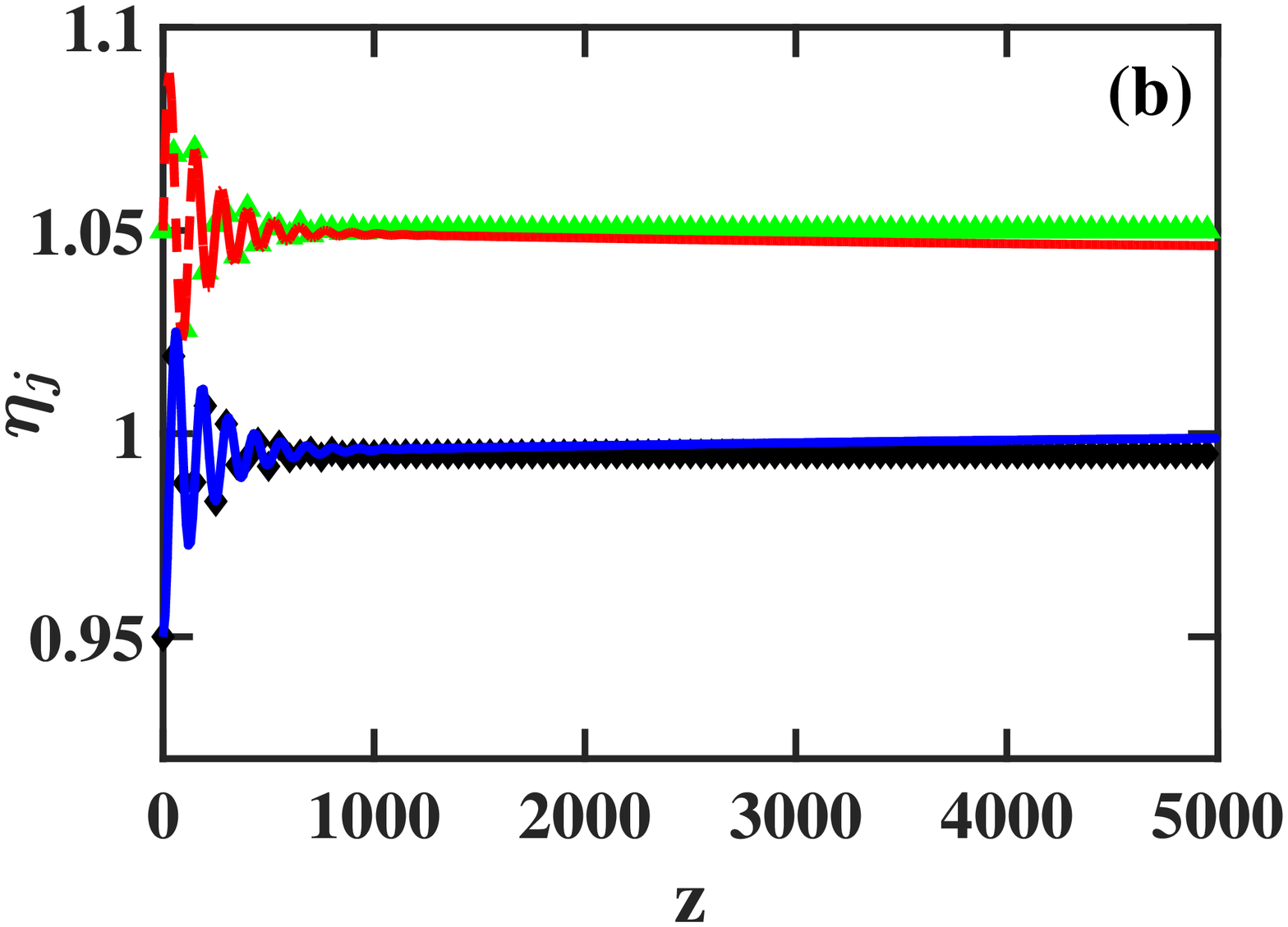}}
\caption{(Color online) The z-dependence of soliton amplitudes $\eta_{j}$ in 
the two-sequence waveguide coupler system of subsection \ref{2D} 
for $\mu=0.98$ (a) and $\mu=1.05$ (b).
The linear gain-loss coefficient is $\epsilon_{1}=0.05$ 
and the initial amplitudes are $\eta_{1}(0)=0.95$ and $\eta_{2}(0)=1.05$.
The solid blue and dashed red curves represent $\eta_{j}(z)$ with $j\!=\!1,2$, 
obtained by numerical simulations with Eq. (\ref{HB1}).  
The black diamonds and green triangles represent $\eta_{j}(z)$ with $j\!=\!1,2$, 
obtained by the LV model (\ref{HB5}).}
\label{fig1}
\end{figure}

\begin{figure}[htbp]
\centerline{\includegraphics[width=.75\columnwidth]{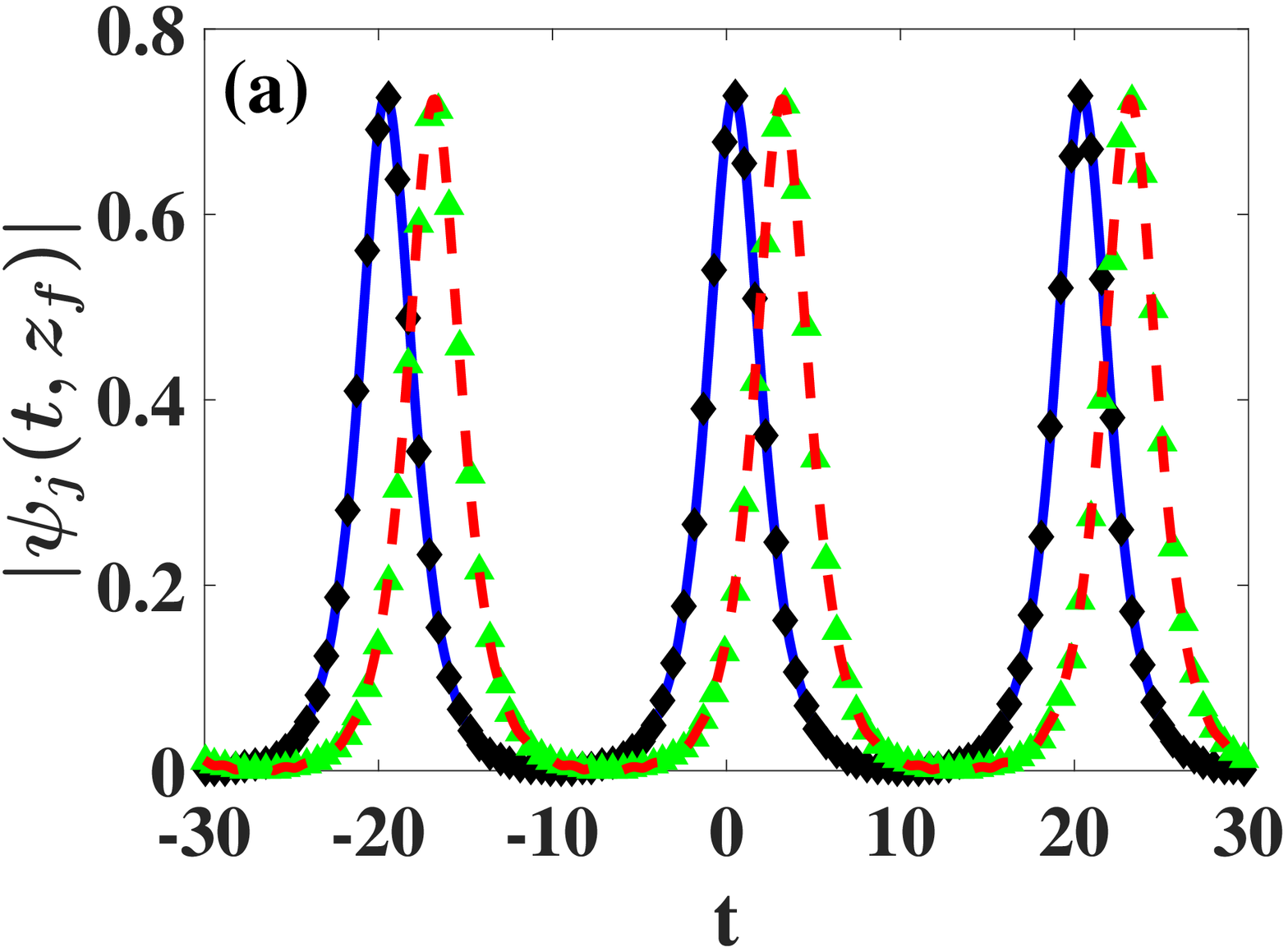}}
\centerline{\includegraphics[width=.75\columnwidth]{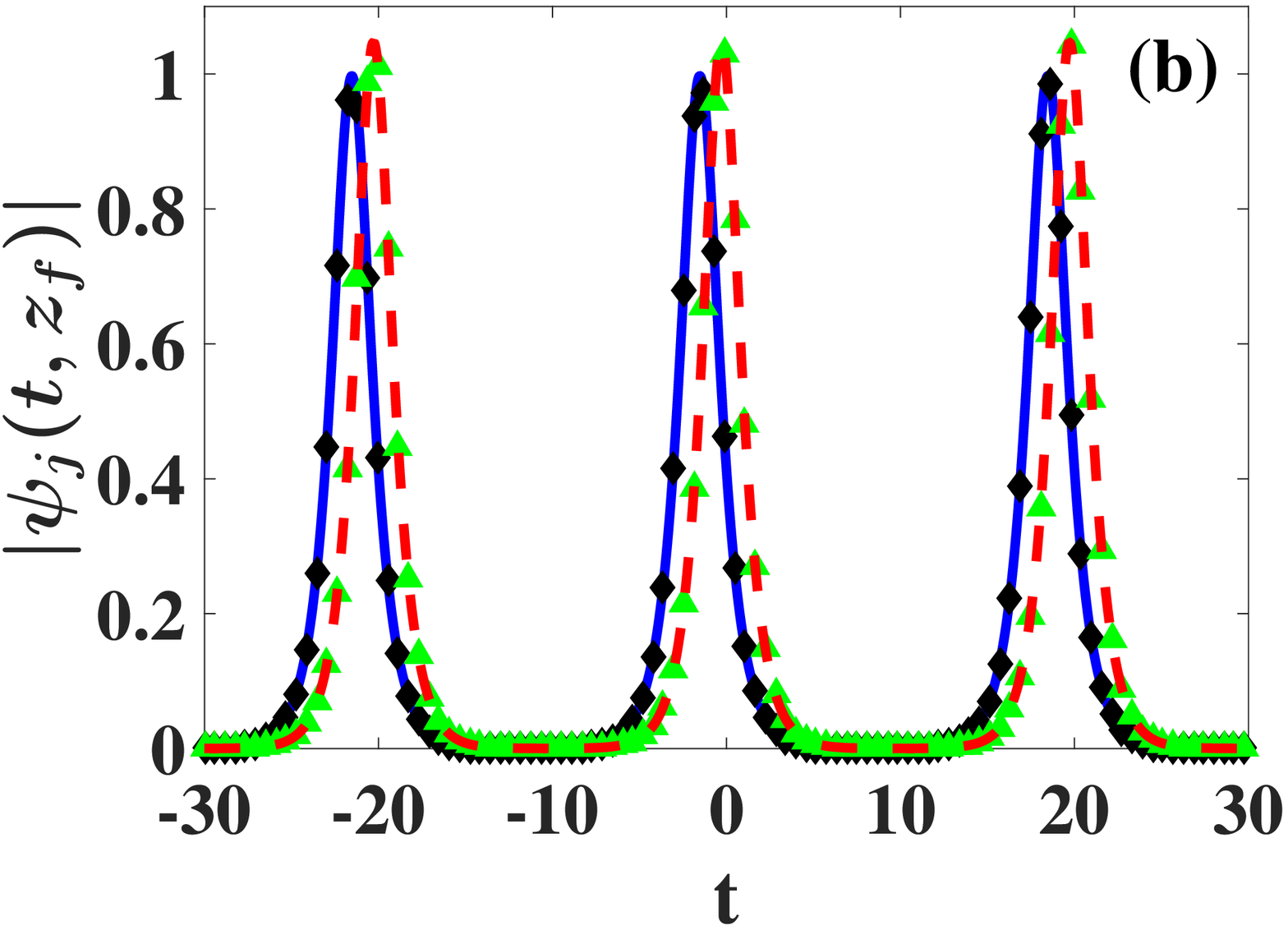}}
\caption{(Color online) The pulse patterns at the final propagation distance $|\psi_{j}(t,z_{f})|$,  
where $z_{f}=5000$, for the two-sequence transmission system of subsection \ref{2D}   
with $\mu=0.98$ (a) and $\mu=1.05$ (b).
The parameter values are the same as in Fig. \ref{fig1}.  
The solid blue and dashed red curves correspond to $|\psi_{j}(t,z_{f})|$
with $j = 1,2$, obtained by numerical simulations with Eq. (\ref{HB1}). 
The black diamonds and green triangles  correspond to the theoretical prediction, 
obtained by summation over fundamental NLS solitons.}
\label{fig2}
\end{figure}

To test these predictions, we numerically solve Eq. (\ref{HB1})
with $\epsilon_{1}=0.05$ for different $\mu$ values close to 1. 
As an example, we present here the results of the simulations with initial soliton 
amplitudes $\eta_{1}(0)=0.95$ and $\eta_{2}(0)=1.05$.  
Figures \ref{fig1}(a) and \ref{fig1}(b) show 
the $z$ dependence of $\eta_{j}$ obtained by the simulations for 
$\mu=0.98$ and $\mu=1.05$, respectively. 
The predictions of the LV model (\ref{HB5}) are also shown. 
It is seen that for $\mu=0.98$, the amplitudes 
exhibit large stable oscillations that approach limit cycle behavior, 
in very good agreement with the LV model's predictions. 
Additionally, for $\mu=1.05$, the amplitudes exhibit decaying oscillations 
and approach their equilibrium values $\eta_{j}^{(eq)}$ in accordance with the LV model.  
Furthermore, as seen in Fig. \ref{fig2}, in both cases 
the solitons retain their shape throughout the propagation.
Similar results are obtained for other values of $\mu$ and the $\eta_{j}(0)$.

\subsection{Three-sequence transmission}
\label{3D}  
\subsubsection{A transmission setup based on the 1980 Arneodo LV model}    
\label{3D_Arneodo} 
We first consider a three-dimensional (3D) LV model for amplitude dynamics in three-sequence waveguide coupler   
transmission, which was studied by Arneodo et al. in Ref. \cite{Arneodo80}. 
The model was introduced and investigated in the context of population dynamics as 
an example for a low-dimensional LV system exhibiting chaos.  
The model is given by Eq. (\ref{HB3}) with  
$N=3$, $g_{11}=1.1$, $g_{21}=-0.5$, $g_{12}=-0.5$, 
$g_{22}=0.1$, $g_{13}=0.2+\mu$, $g_{23}=-0.1$,  
$g_{31}=g_{32}=g_{33}=0$, 
$\epsilon_{312}=0.5\epsilon_{1}T/8$, $\epsilon_{313}=0.1\epsilon_{1}T/8$,       
$\epsilon_{321}=-0.5\epsilon_{1}T/8$, $\epsilon_{323}=0.1\epsilon_{1}T/8$, 
$\epsilon_{331}=\mu\epsilon_{1}T/8$, $\epsilon_{332}=0.1\epsilon_{1}T/8$.
Thus, this LV model is:                 
\begin{eqnarray} &&  
\frac{d\eta _{1}}{dz}=
\epsilon_{1}\eta_{1}\left(1.1 - 0.5\eta_{1} - 0.5\eta_{2} - 0.1\eta_{3}\right), 
\nonumber \\&&
\frac{d\eta _{2}}{dz}=
\epsilon_{1}\eta_{2}\left(-0.5 + 0.5\eta_{1} + 0.1\eta_{2} - 0.1\eta_{3}\right),
\nonumber \\&&
\frac{d\eta _{3}}{dz}=
\epsilon_{1}\eta_{3}\left(0.2 +\mu - \mu\eta_{1} - 0.1\eta_{2} - 0.1\eta_{3}\right). 
\label{HB6}
\end{eqnarray}                                  
Note that in this waveguide coupler system, sequences $j=1$ and $j=3$ 
propagate in the presence of cubic loss intersequence interaction,
while the $j=2$ sequence propagates in the presence of cubic gain 
intersequence interaction with sequence $j=1$ and cubic loss 
intersequence interaction with sequence $j=3$.           
The relevant equilibrium state of the LV model is $(1,1,1)$, independent of 
the value of $\mu$ \cite{Arneodo80}. This equilibrium state undergoes a supercritical Hopf bifurcation 
as  $\mu$ increases through $\mu_{H} \simeq 0.954$ \cite{Arneodo80}. 
As a result, for $\mu<\mu_{H}$, the equilibrium state is a stable focus, 
while for $\mu>\mu_{H}$, $(1,1,1)$ becomes unstable and a stable limit cycle 
about $(1,1,1)$ appears. As $\mu$ increases beyond $\mu_{P} \simeq 1.265$, 
the limit cycle undergoes a period doubling cascade, 
and finally, chaotic dynamics is observed \cite{Arneodo80}.       
Thus, we expect the amplitudes to exhibit stable decaying oscillations 
and approach the equilibrium value 1 for $\mu<\mu_{H}$, and to exhibit large stable oscillations 
with a single period for $\mu_{H}<\mu<\mu_{P}$.

\begin{figure}[htbp]
\centerline{\includegraphics[width=.75\columnwidth]{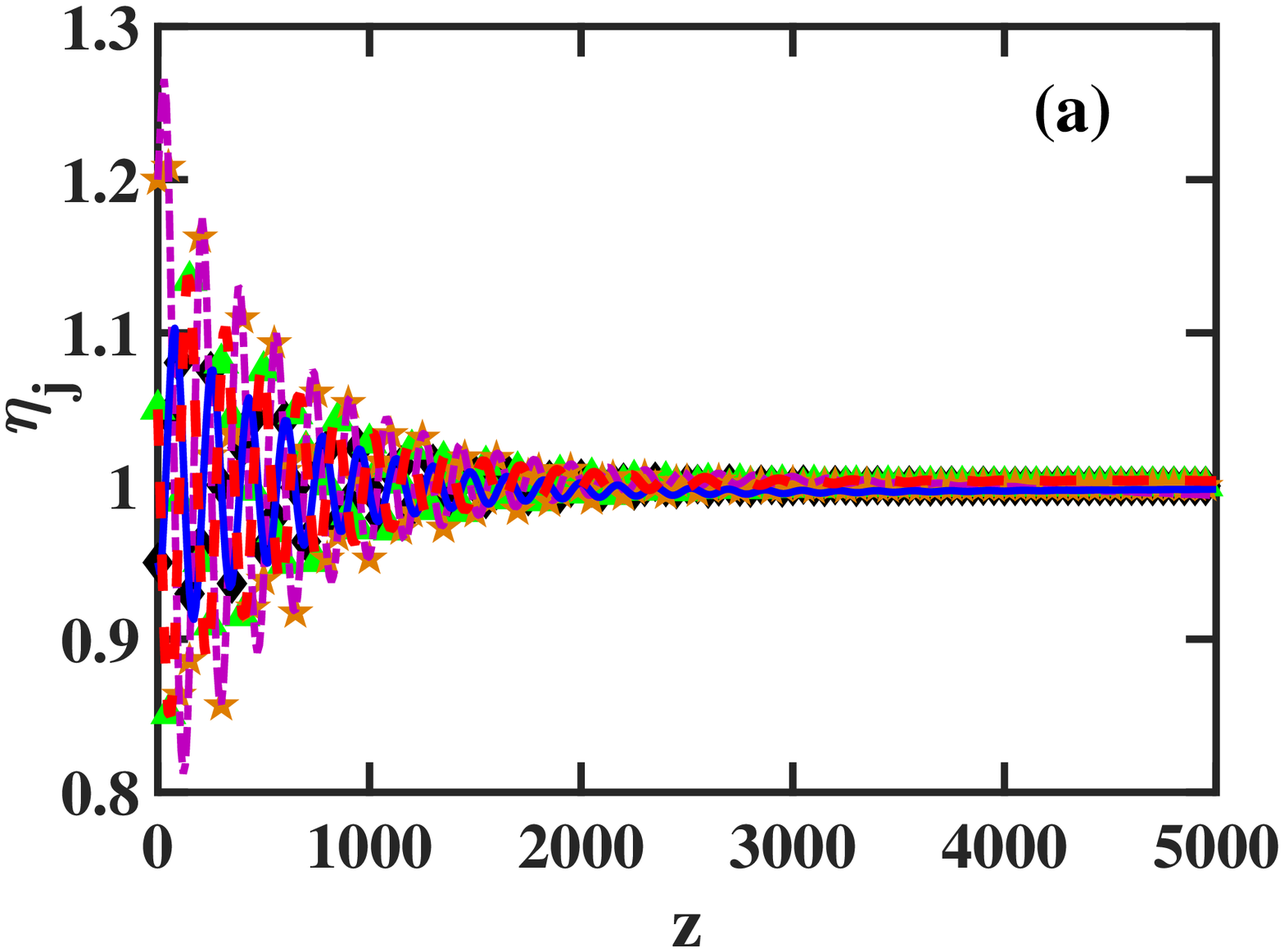}}
\centerline{\includegraphics[width=.75\columnwidth]{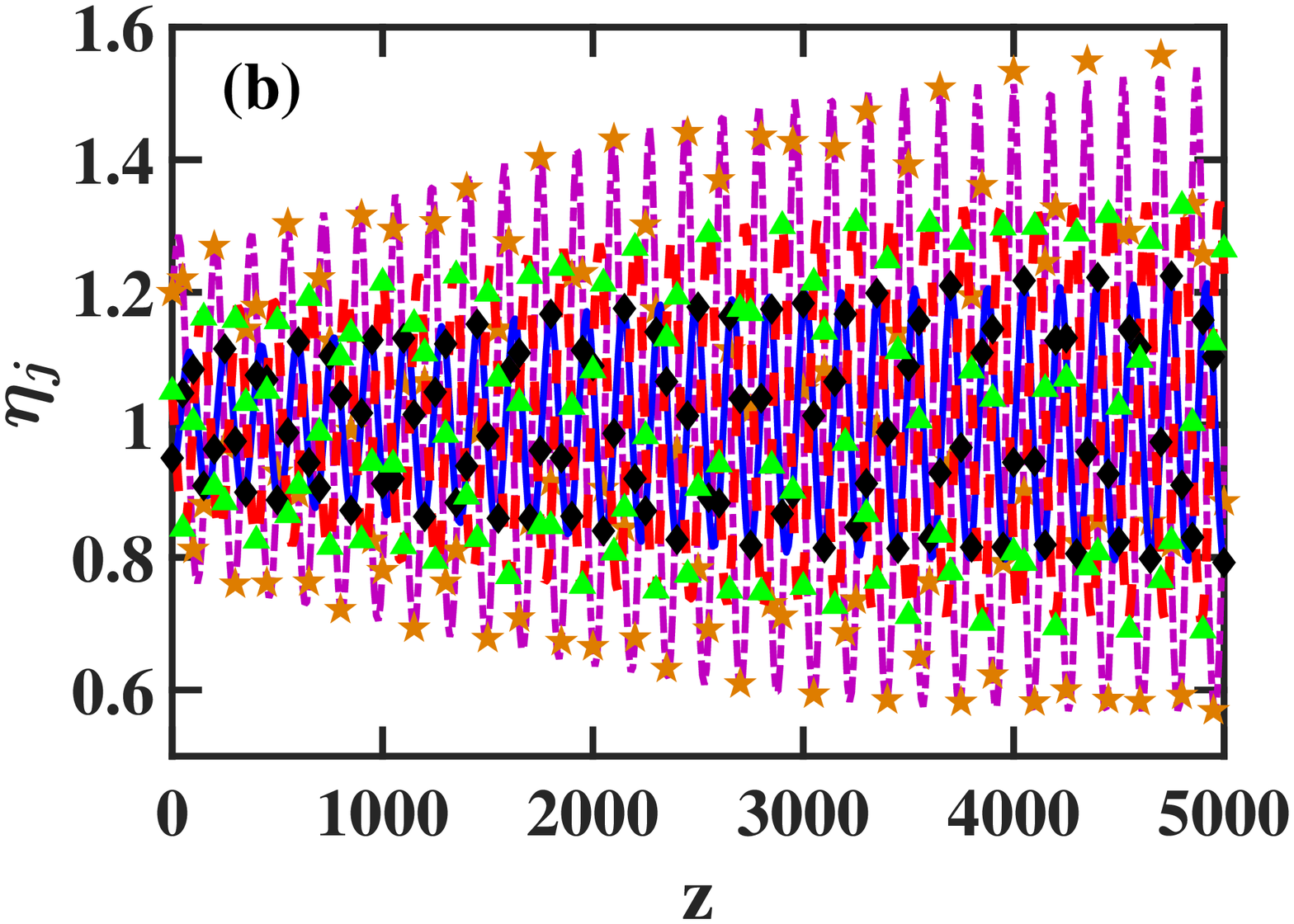}}
\caption{(Color online) The z-dependence of soliton amplitudes $\eta_{j}$ 
in the three-sequence waveguide coupler system of subsection \ref{3D_Arneodo}  
for $\mu=0.85$ (a) and $\mu=0.98$ (b).
The linear gain-loss coefficient is $\epsilon_{1}=0.1$ 
and the initial amplitudes are $\eta_{1}(0)=0.95$, $\eta_{2}(0)=1.05$, and $\eta_{3}(0)=1.2$.
The solid blue, dashed red, and dash-dotted purple curves represent 
$\eta_{j}(z)$ with $j=1,2,3$, obtained by numerical simulations with Eq. (\ref{HB1}).  
The black diamonds, green triangles, and orange stars represent $\eta_{j}(z)$ with $j=1,2,3$, 
obtained by the LV model (\ref{HB6}).} 
\label{fig3}
\end{figure}

\begin{figure}[htbp]
\centerline{\includegraphics[width=.75\columnwidth]{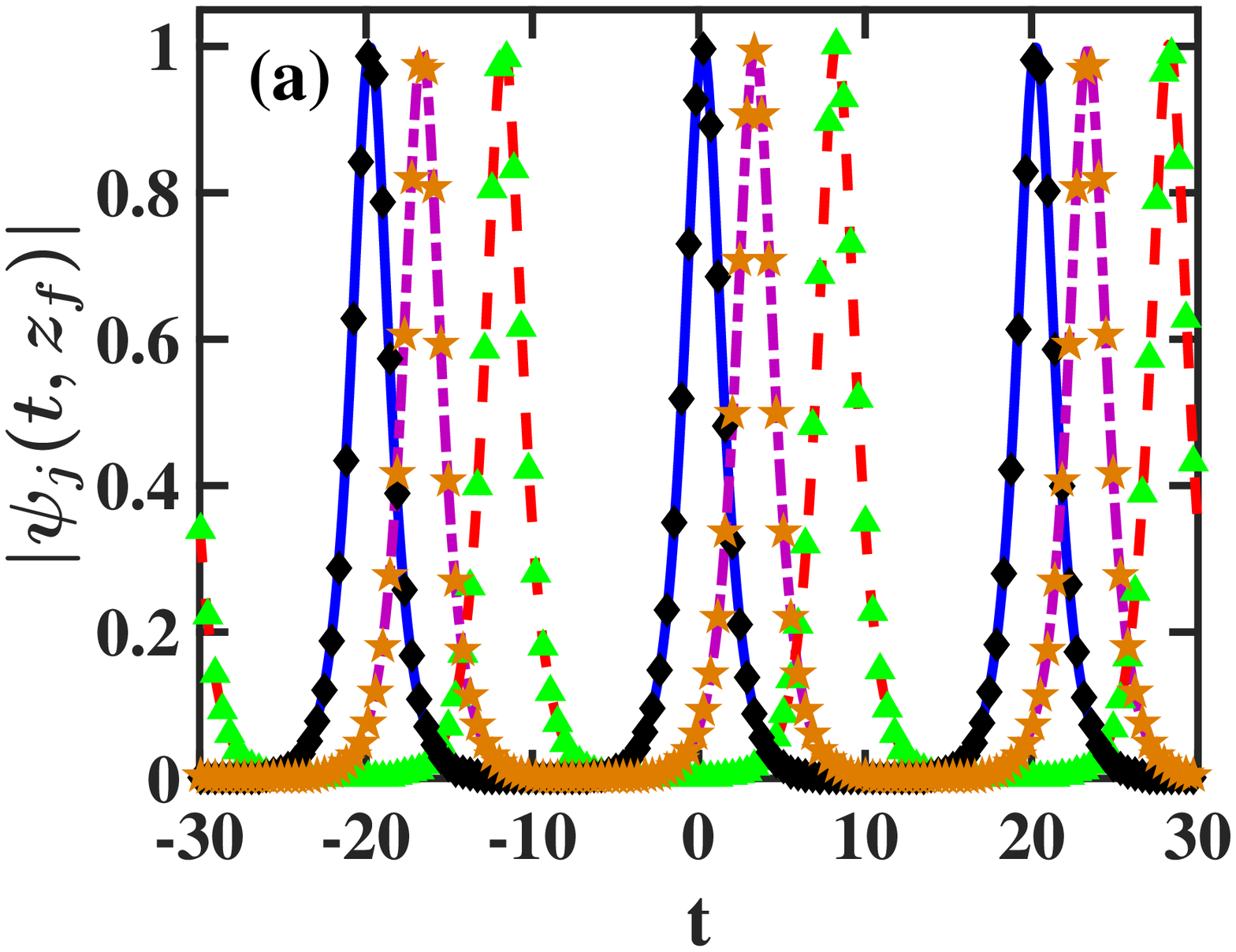}}
\centerline{\includegraphics[width=.75\columnwidth]{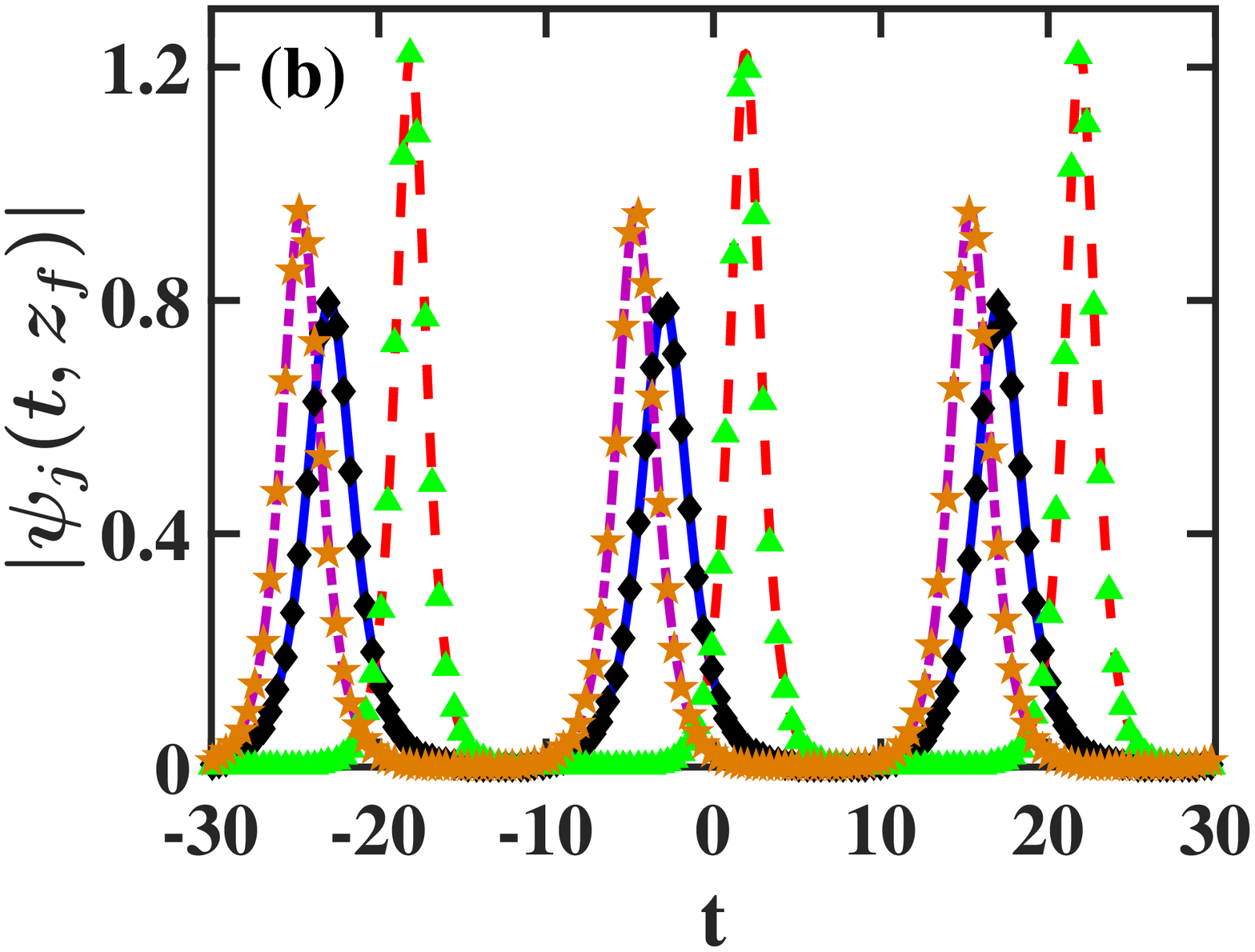}}
\caption{(Color online) The pulse patterns at the final propagation distance $|\psi_{j}(t,z_{f})|$,  
where $z_{f}=5000$, for the three-sequence transmission 
system of subsection \ref{3D_Arneodo} with $\mu=0.85$ (a) 
and $\mu=0.98$ (b). The parameter values are the same as in Fig. \ref{fig3}.   
The solid blue, dashed red, and dash-dotted purple curves 
correspond to $|\psi_{j}(t,z_{f})|$ with $j = 1,2,3$, obtained by numerical simulations with Eq. (\ref{HB1}). 
The black diamonds, green triangles, and orange stars correspond to the theoretical prediction, 
obtained by summation over fundamental NLS solitons.}
\label{fig4}
\end{figure}

To validate the predictions of the LV model, we numerically solve Eq. (\ref{HB1}) 
with $\epsilon_{1}=0.1$ for different $\mu$ values. 
As an example, we present here the results of the simulations with initial  
amplitudes  $\eta_{1}(0)=0.95$, $\eta_{2}(0)=1.05$, and $\eta_{3}(0)=1.2$. 
Figures \ref{fig3}(a) and \ref{fig3}(b) show 
the $z$ dependence of $\eta_{j}$ obtained by the simulations for 
$\mu=0.85$ and $\mu=0.98$, respectively, 
together with the predictions of the LV model (\ref{HB6}). 
The agreement between the numerical simulations and the predictions of 
the LV model is very good for both values of $\mu$. 
In particular, for $\mu=0.85$, the amplitudes exhibit decaying oscillations 
and approach their equilibrium value of 1, while for $\mu=0.98$, 
the amplitudes exhibit large stable oscillations that tend to limit cycle behavior. 
Furthermore, as seen in Fig. \ref{fig4},  
in both cases the soliton patterns remain intact throughout the propagation. 
Similar results are obtained for other values of $\mu$ and the $\eta_{j}(0)$.

\subsubsection{A simpler three-sequence transmission setup}    
\label{3D_Odell}  
We now consider a second example for a 3D LV model 
for amplitude dynamics in three-sequence waveguide coupler   
transmission, for which the soliton amplitudes exhibit large stable 
oscillations due to a supercritical Hopf bifurcation. This 3D LV model does not exhibit 
chaotic dynamics, but it has the advantage of having a simpler form compared 
with the 3D model of Eq. (\ref{HB6}). The model is given by Eq. (\ref{HB3}) with $N=3$, 
$g_{11}=-2\mu$, $g_{21}=0$, $g_{12}=0$, 
$g_{22}=-1$, $g_{13}=2$, $g_{23}=-1$,  
$g_{31}=g_{32}=g_{33}=0$, 
$\epsilon_{312}=0$, $\epsilon_{313}=-\epsilon_{1}T/4$,       
$\epsilon_{321}=-\epsilon_{1}T/8$, $\epsilon_{323}=0$, 
$\epsilon_{331}=0$, $\epsilon_{332}=\epsilon_{1}T/8$.
Therefore, the LV model is:                 
\begin{eqnarray} &&  
\frac{d\eta _{1}}{dz}=
2\epsilon_{1}\eta_{1}\left(-\mu+\eta_{3}\right), 
\nonumber \\&&
\frac{d\eta _{2}}{dz}=
\epsilon_{1}\eta_{2}\left(\eta_{1}-\eta_{2}\right), 
\nonumber \\&&
\frac{d\eta _{3}}{dz}=
\epsilon_{1}\eta_{3}\left(2-\eta_{2}-\eta_{3}\right). 
\label{HB7}
\end{eqnarray}                             
In this waveguide coupler system, sequences $j=1$ and $j=2$ 
propagate in the presence of cubic gain intersequence interaction,
while the $j=3$ sequence propagates in the presence of cubic loss 
intersequence interaction. The relevant equilibrium state is 
$(\eta_{1}^{(eq)},\eta_{2}^{(eq)},\eta_{3}^{(eq)})\!=\!(2-\mu,2-\mu,\mu)$. 
Stability analysis shows that this equilibrium state undergoes 
a supercritical Hopf bifurcation as  $\mu$ changes through 1. 
For $0<\mu<1$, the equilibrium state is an unstable focus, 
which is surrounded by a stable limit cycle. Additionally, for $1<\mu<2$, 
it is a stable focus and the limit cycle does not exist.           
Therefore, we expect large stable amplitude oscillations 
for $0<\mu<1$, and stable oscillations that decay to 
$\eta_{j}^{(eq)}$ for $1<\mu<2$.

\begin{figure}[htbp]
\centerline{\includegraphics[width=.75\columnwidth]{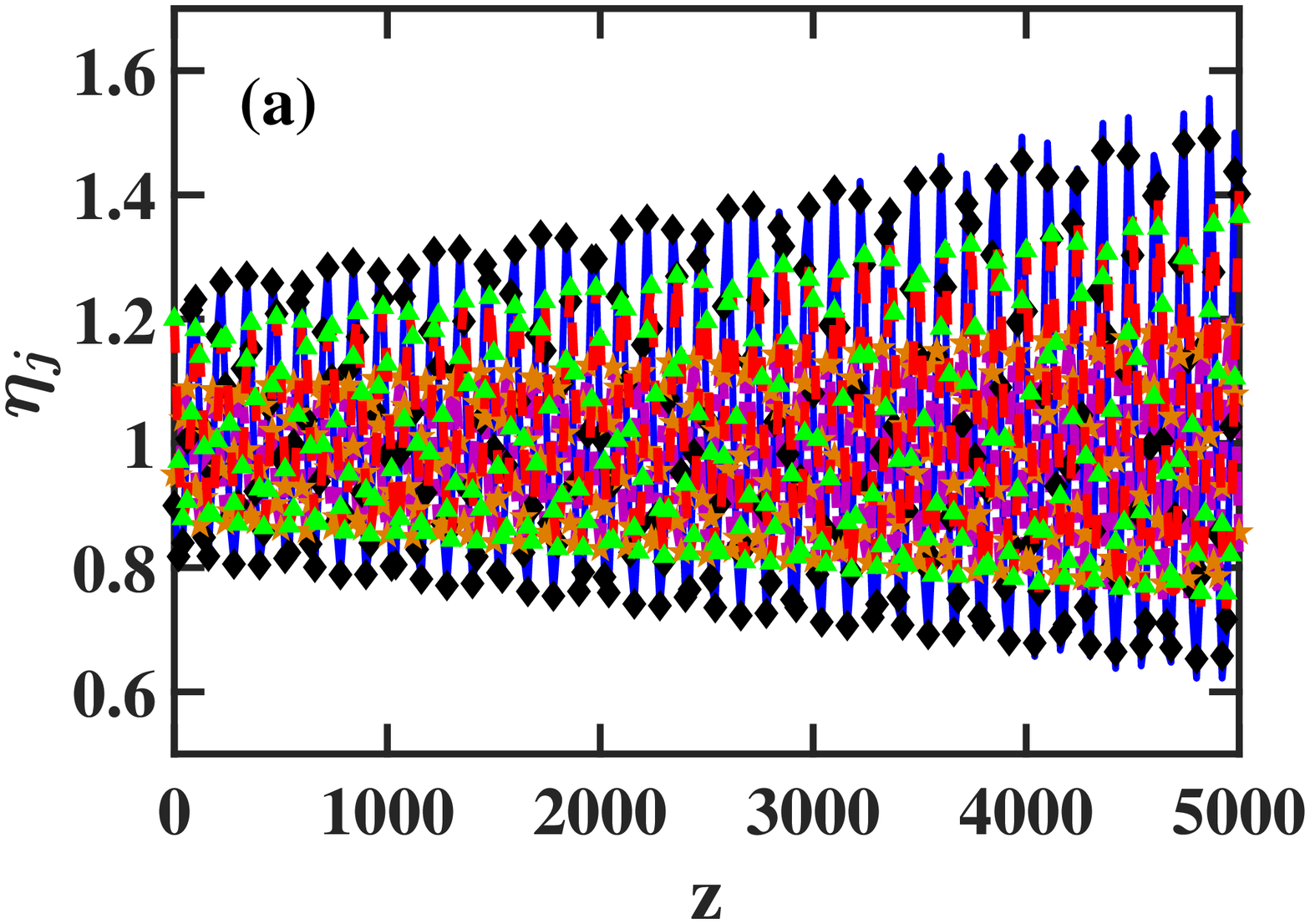}}
\centerline{\includegraphics[width=.75\columnwidth]{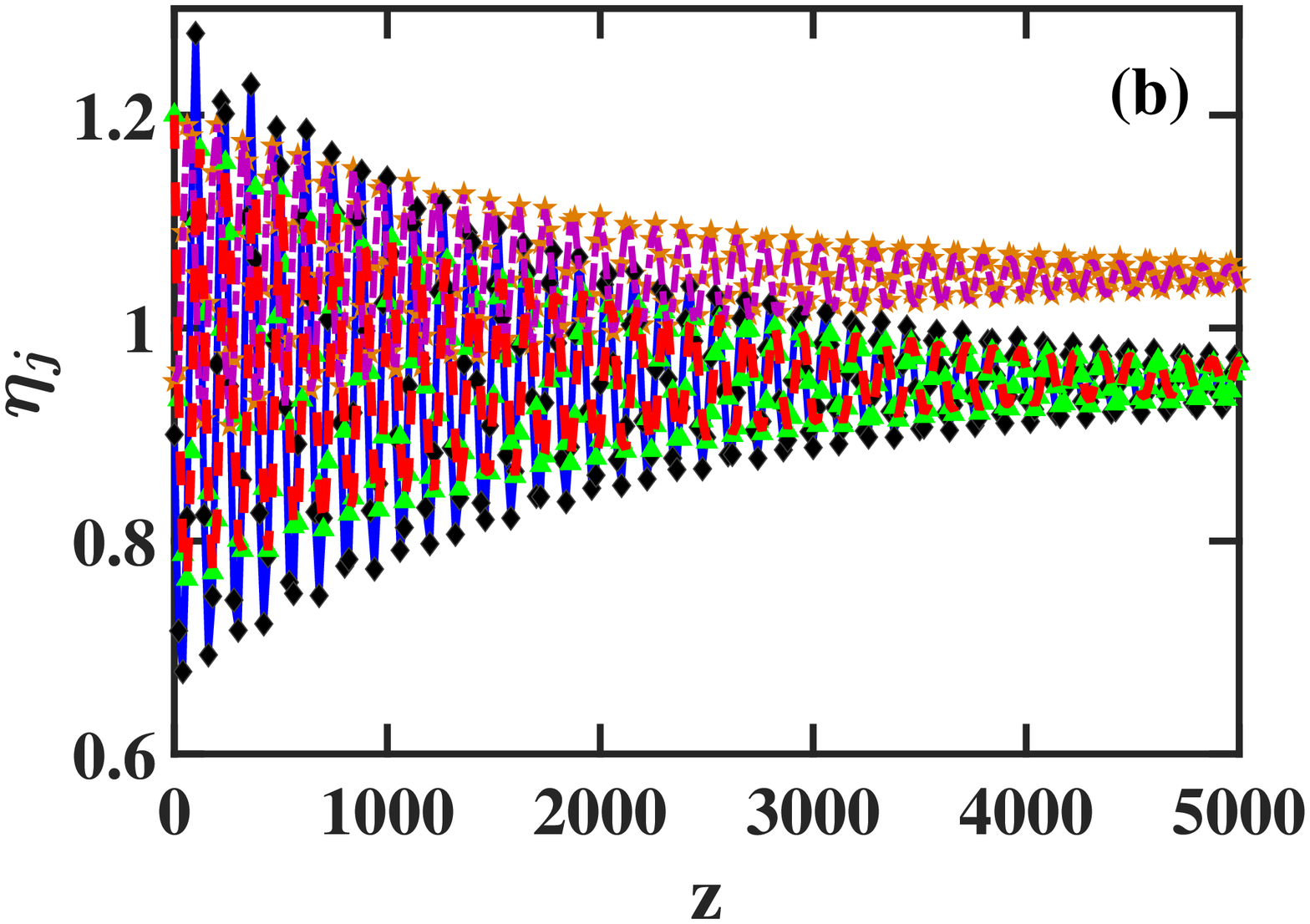}}
\caption{(Color online) The z-dependence of soliton amplitudes $\eta_{j}$ 
in the three-sequence waveguide coupler system of subsection \ref{3D_Odell}    
for $\mu=0.98$ (a) and $\mu=1.05$ (b).
The linear gain-loss coefficient is $\epsilon_{1}=0.05$ 
and the initial amplitudes are $\eta_{1}(0)=0.9$, $\eta_{2}(0)=1.2$, and $\eta_{3}(0)=0.95$. 
The solid blue, dashed red, and dash-dotted purple curves represent 
$\eta_{j}(z)$ with $j=1,2,3$, obtained by numerical simulations with Eq. (\ref{HB1}).  
The black diamonds, green triangles, and orange stars represent $\eta_{j}(z)$ with $j=1,2,3$, 
obtained by the LV model (\ref{HB7}).} 
\label{add_fig1}
\end{figure}

\begin{figure}[htbp]
\centerline{\includegraphics[width=.75\columnwidth]{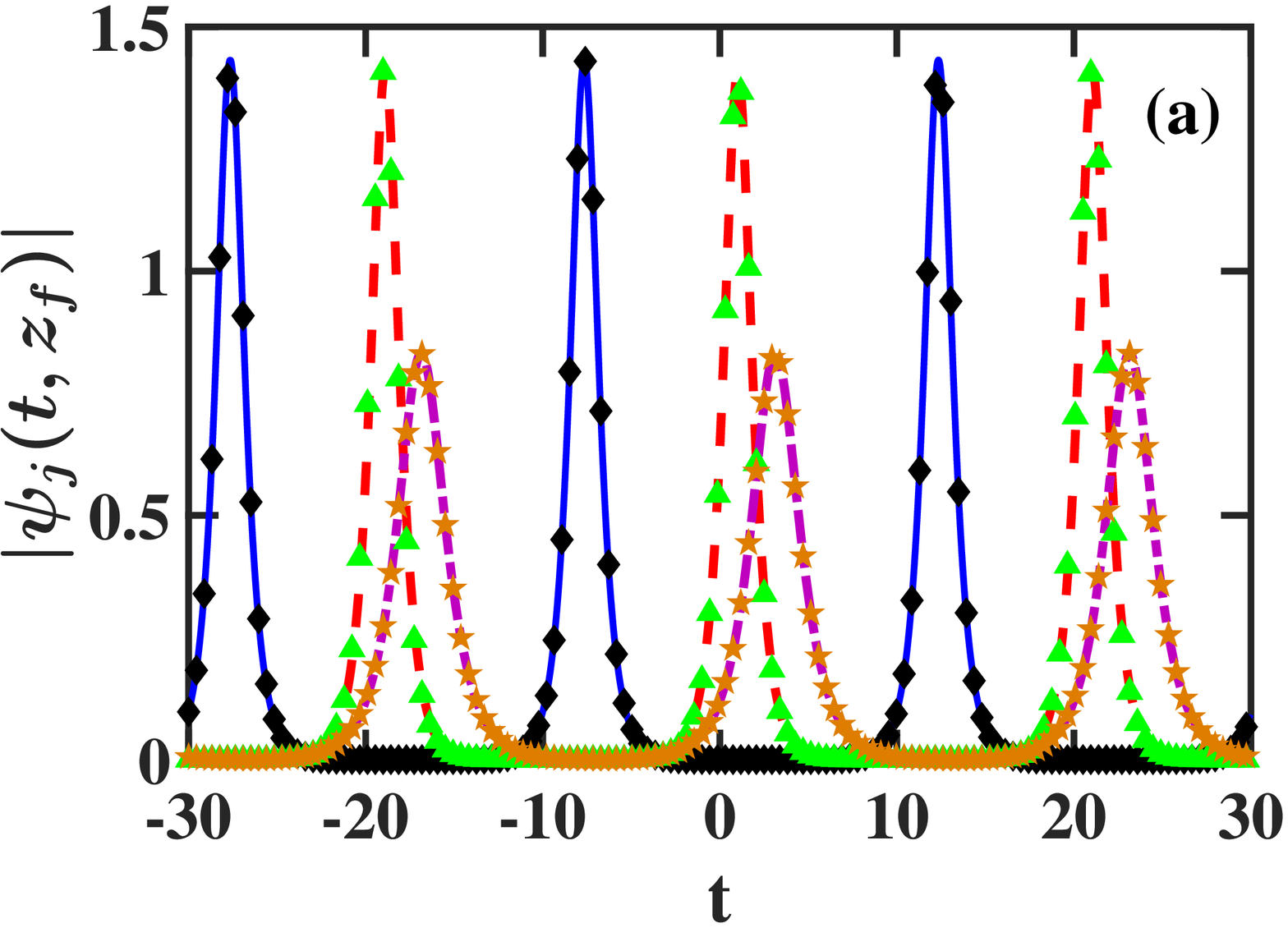}}
\centerline{\includegraphics[width=.75\columnwidth]{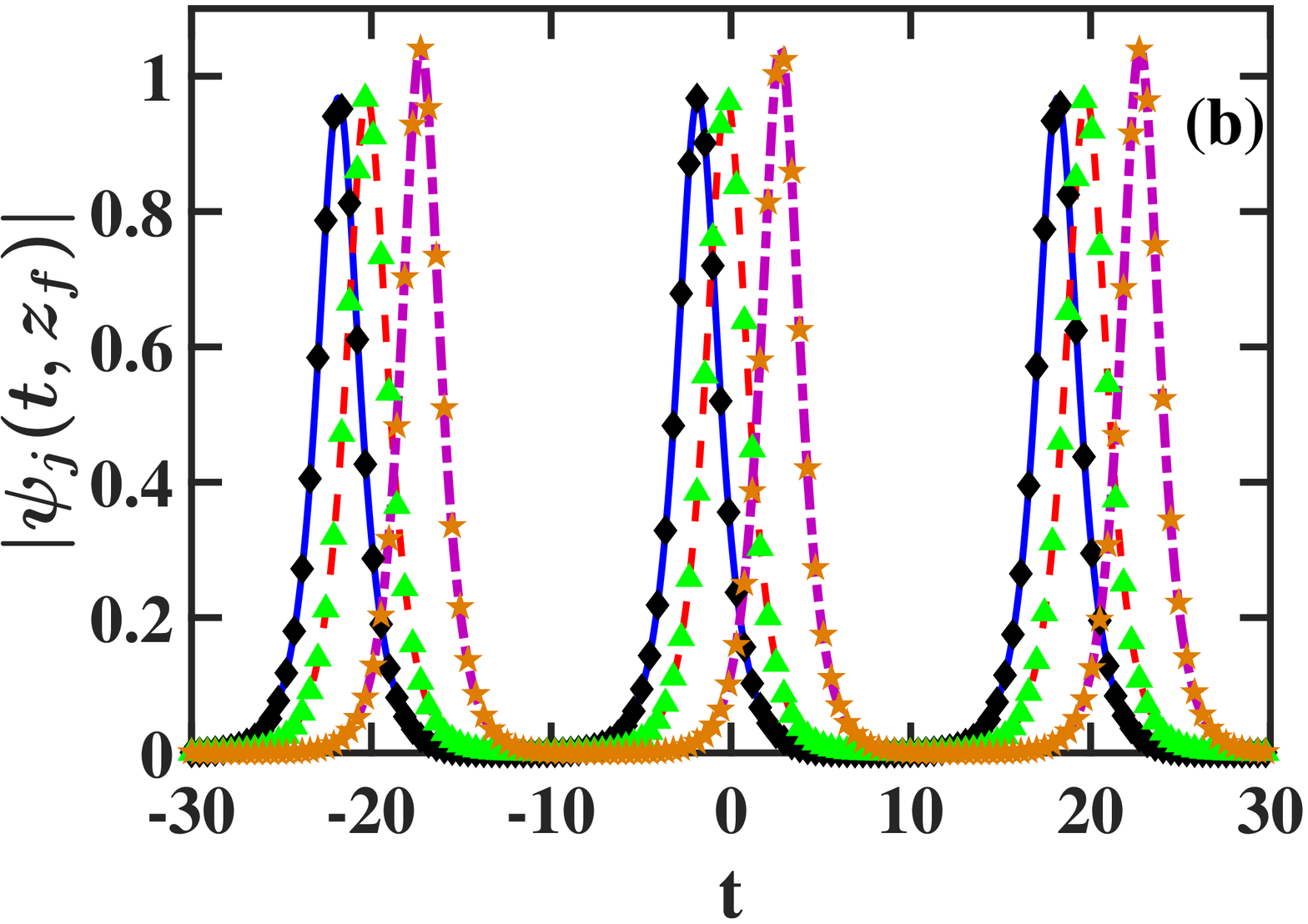}}
\caption{(Color online) The pulse patterns at the final propagation distance $|\psi_{j}(t,z_{f})|$,  
where $z_{f}=5000$, for the three-sequence transmission system of subsection \ref{3D_Odell}    
with $\mu=0.98$ (a) and $\mu=1.05$ (b). The parameter values are the same as in Fig. \ref{add_fig1}.   
The solid blue, dashed red, and dash-dotted purple curves 
correspond to $|\psi_{j}(t,z_{f})|$ with $j = 1,2,3$, obtained by numerical simulations with Eq. (\ref{HB1}). 
The black diamonds, green triangles, and orange stars correspond to the theoretical prediction, 
obtained by summation over fundamental NLS solitons.}
\label{add_fig2}
\end{figure}

To check the predictions of the LV model, we numerically solve Eq. (\ref{HB1}) 
with $\epsilon_{1}=0.05$ for different $\mu$ values. 
As an example, we present here the results of the simulations with initial  
soliton amplitudes $\eta_{1}(0)=0.9$, $\eta_{2}(0)=1.2$, and $\eta_{3}(0)=0.95$. 
Figures \ref{add_fig1}(a) and \ref{add_fig1}(b) show 
the $z$ dependence of $\eta_{j}$ obtained by the simulations for 
$\mu=0.98$ and $\mu=1.05$, respectively,
along with the predictions of the LV model (\ref{HB7}). 
The agreement between the numerical simulations and the predictions of 
the LV model is very good for both values of $\mu$. 
In particular, for $\mu=0.98$, the amplitudes exhibit large stable oscillations 
that tend to limit cycle behavior, while for $\mu=1.05$, the amplitudes exhibit 
decaying oscillations and approach their equilibrium values $\eta_{j}^{(eq)}$.
Furthermore, as seen in Fig. \ref{add_fig2},  
for both values of $\mu$, the solitons retain their shape throughout the propagation. 
Similar results are obtained for other values of $\mu$ and the $\eta_{j}(0)$.

\subsection{Four-sequence transmission}
\label{4D}              
The four-dimensional (4D) LV model for four-sequence waveguide coupler transmission 
was studied in Ref. \cite{Arneodo82} in the context of population dynamics    
by Arneodo et al. This LV model is given by Eq. (\ref{HB3}) with   
$N=4$, $g_{1j}=3.3$ for $j=1,2,3,4$, $g_{21}=-1$, 
$g_{22}=-0.4$, $g_{23}=-0.6$, $g_{24}=-1.8$, 
$g_{3j}=0$ for $j=1,2,3,4$, 
$\epsilon_{312}=\epsilon_{1}T/8$, $\epsilon_{313}=0.6\epsilon_{1}T/8$, $\epsilon_{314}=0.7\epsilon_{1}T/8$,      
$\epsilon_{321}=0$, $\epsilon_{323}=0.6\epsilon_{1}T/8$, $\epsilon_{324}=2.3\epsilon_{1}T/8$, 
$\epsilon_{331}=(\mu+0.5)\epsilon_{1}T/8$, $\epsilon_{332}=0.6\epsilon_{1}T/8$, 
$\epsilon_{334}=(1.6-\mu)\epsilon_{1}T/8$, and 
$\epsilon_{34j}=0.5\epsilon_{1}T/8$ for $j=1,2,3,4$.   
Therefore, this LV model is:                 
\begin{eqnarray} &&  
\frac{d\eta _{1}}{dz}=
\epsilon_{1}\eta_{1}\left(3.3 - \eta_{1} - \eta_{2} - 0.6\eta_{3} - 0.7\eta_{4}\right), 
\nonumber \\&&
\frac{d\eta _{2}}{dz}=
\epsilon_{1}\eta_{2}\left(3.3 - 0.4\eta_{2} - 0.6\eta_{3} - 2.3\eta_{4}\right),
\nonumber \\&&
\frac{d\eta _{3}}{dz}=
\epsilon_{1}\eta_{3}\left[3.3 - (\mu+0.5)\eta_{1} - 0.6\eta_{2} - 0.6\eta_{3} -  (1.6-\mu)\eta_{1} \right],
\nonumber \\&&
\frac{d\eta _{4}}{dz}=
\epsilon_{1}\eta_{4}\left(3.3 - 0.5\eta_{1} - 0.5\eta_{2} - 0.5\eta_{3} - 1.8\eta_{4}\right).
\label{HB8}
\end{eqnarray}                                 
Note that in this transmission system, all soliton sequences propagate in the presence of cubic loss 
intersequence interaction. This is very different from the two-sequence and 
three-sequence systems considered in the preceding subsections, 
where at least one of the soliton sequences propagated in the presence of cubic gain interaction. 
The 4D LV model (\ref{HB8}) is in fact equivalent to the 3D LV (\ref{HB6}) (see Ref. \cite{Coste79}). 
Therefore, the relevant equilibrium state for amplitude dynamics is $(1,1,1,1)$. 
This equilibrium state undergoes a supercritical Hopf bifurcation 
at $\mu_{H} \simeq 0.954$ \cite{Arneodo82}, and as a result, 
the state $(1,1,1,1)$ turns from a stable focus to an unstable state 
and a stable limit cycle about $(1,1,1,1)$ appears. As $\mu$ is 
increased beyond $\mu_{P} \simeq 1.265$, the limit cycle undergoes a period doubling cascade, 
and finally, chaotic dynamics is observed \cite{Arneodo82}.                  
Thus, we expect the soliton amplitudes to exhibit stable decaying oscillations 
and approach 1 for $\mu<\mu_{H}$, and to exhibit large stable oscillations 
with a single period for $\mu_{H}<\mu<\mu_{P}$.

\begin{figure}[htbp]
\centerline{\includegraphics[width=.75\columnwidth]{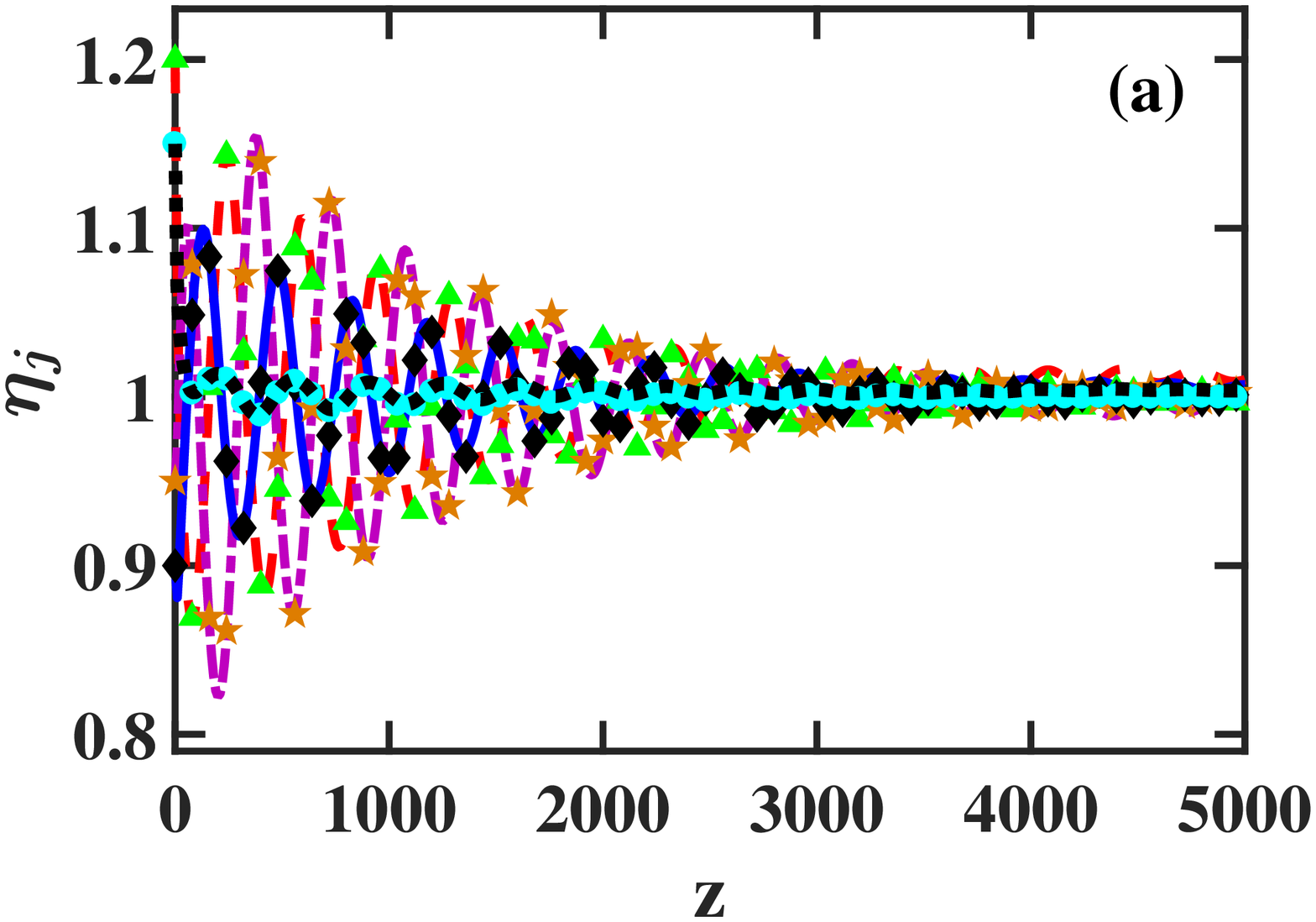}}
\centerline{\includegraphics[width=.75\columnwidth]{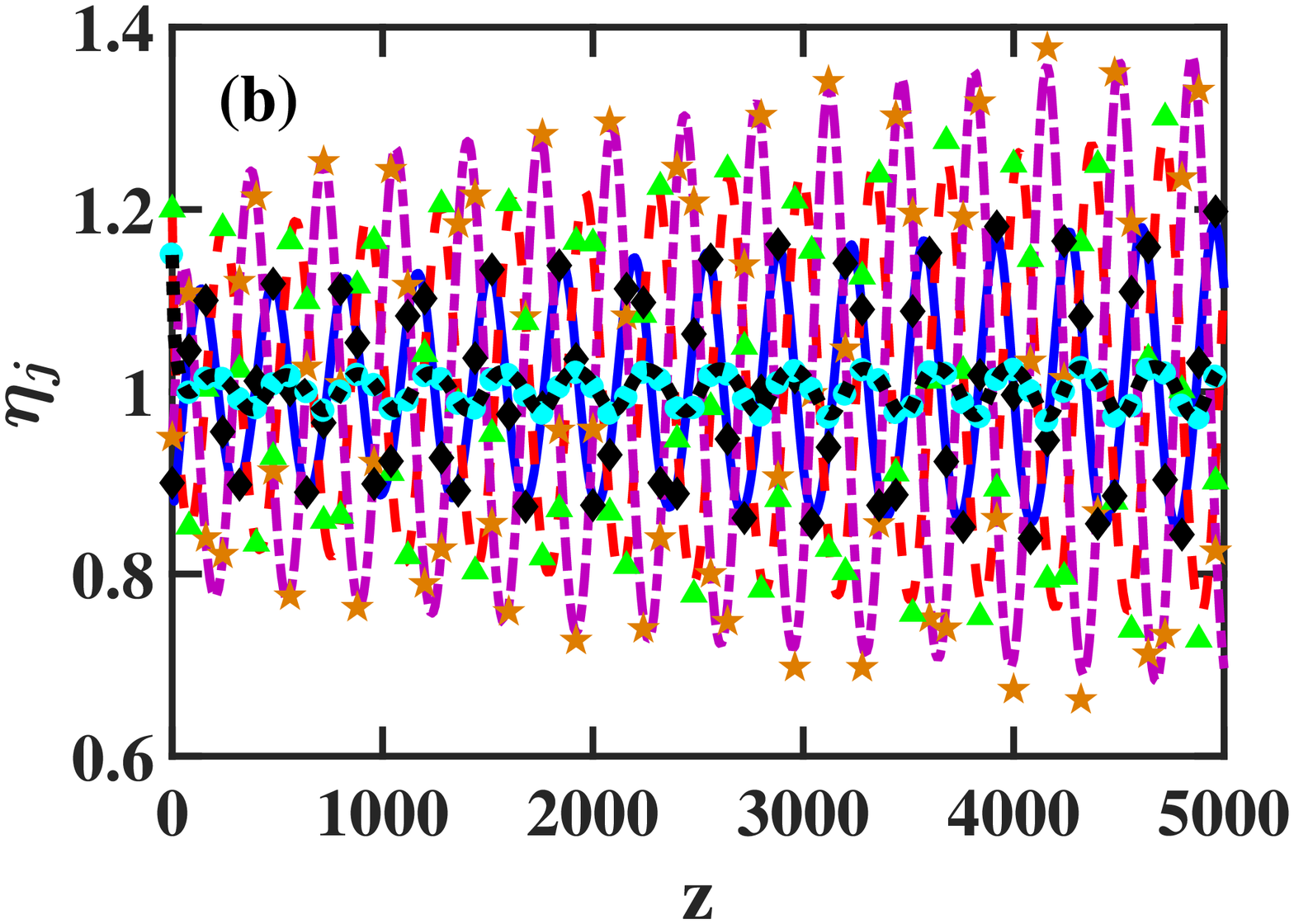}}
\caption{(Color online) Soliton amplitudes $\eta_{j}$ vs $z$  
in the four-sequence waveguide coupler system of subsection \ref{4D} 
for $\mu=0.85$ (a) and $\mu=0.98$ (b).
The linear gain-loss coefficient is $\epsilon_{1}=0.05$ and the initial amplitudes are 
$\eta_{1}(0) = 0.9$, $\eta_{2}(0) = 1.2$, $\eta_{3}(0) = 0.95$, and $\eta_{4}(0) = 1.15$.
The solid blue, dashed red, dash-dotted purple, and dotted black curves represent $\eta_{j}(z)$ 
with $j=1,2,3,4$, obtained by numerical simulations with Eq. (\ref{HB1}).  
The black diamonds, green triangles, orange stars, and cyan circles
represent $\eta_{j}(z)$ with $j=1,2,3,4$, obtained by the LV model (\ref{HB8}).}
\label{fig5}
\end{figure}

\begin{figure}[htbp]
\centerline{\includegraphics[width=.75\columnwidth]{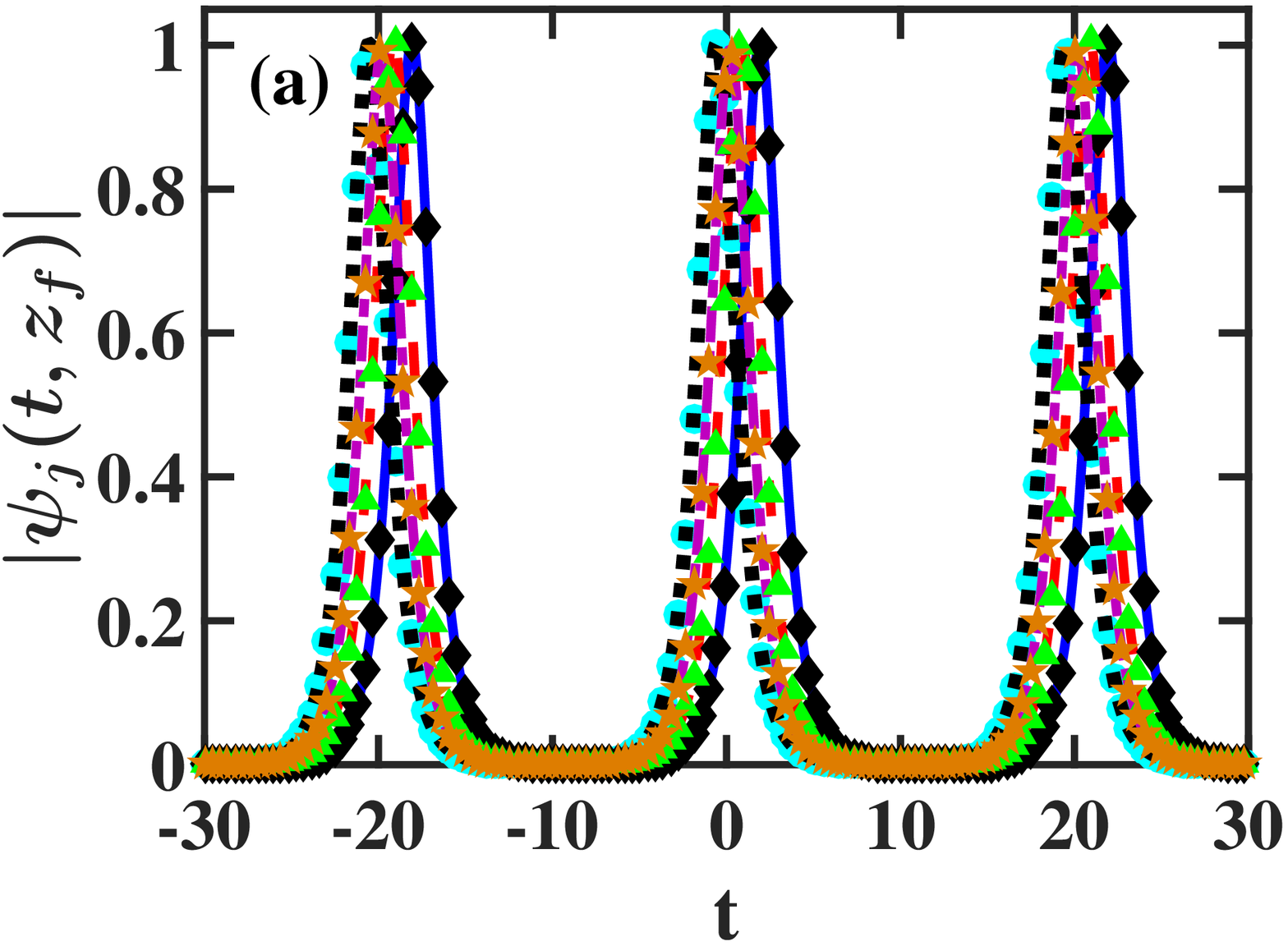}}
\centerline{\includegraphics[width=.75\columnwidth]{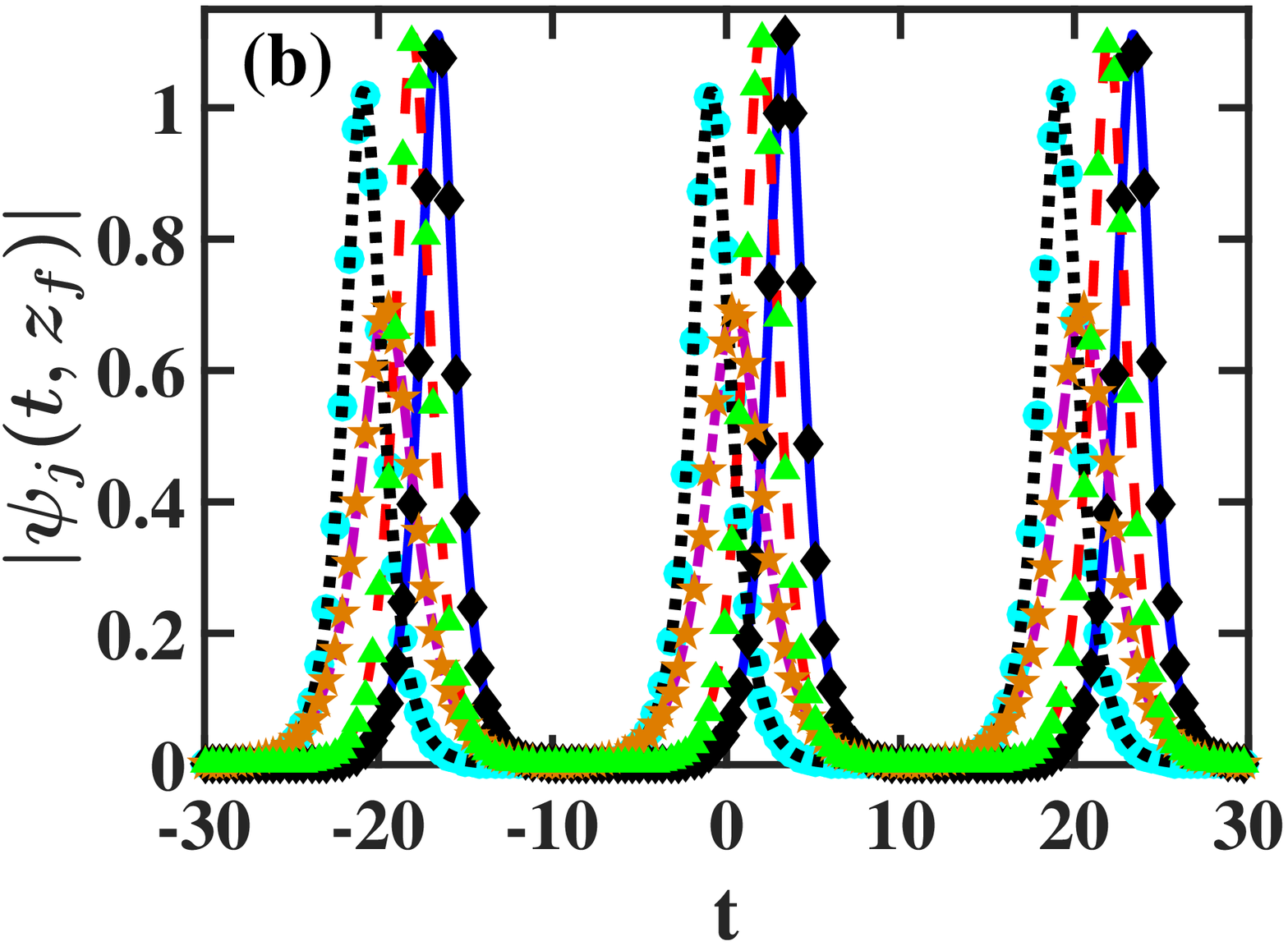}}
\caption{(Color online) The pulse patterns at the final propagation distance $|\psi_{j}(t,z_{f})|$,  
where $z_{f}=5000$, for the four-sequence transmission system of subsection \ref{4D} with $\mu=0.85$ (a) 
and $\mu=0.98$ (b). The parameter values are the same as in Fig. \ref{fig5}.    
The solid blue, dashed red, dash-dotted purple, and dotted black curves 
correspond to $|\psi_{j}(t,z_{f})|$ with $j = 1,2,3,4$, obtained by numerical simulations with Eq. (\ref{HB1}). 
The black diamonds, green triangles, orange stars, and cyan circles correspond 
to the theoretical prediction, obtained by summation over fundamental NLS solitons.}
\label{fig6}
\end{figure}

To check the predictions of the 4D LV model, we numerically solve Eq. (\ref{HB1}) 
with $\epsilon_{1}=0.05$ for different $\mu$ values. 
As an example, we present here the results of the simulations with initial  
pulse amplitudes $\eta_{1}(0) = 0.9$, $\eta_{2}(0) = 1.2$, $\eta_{3}(0) = 0.95$, 
and $\eta_{4}(0) = 1.15$.   
Figures \ref{fig5}(a) and \ref{fig5}(b) show the $z$ dependence 
of $\eta_{j}$ obtained by the simulations for $\mu=0.85$ and $\mu=0.98$, respectively. 
The predictions of the LV model (\ref{HB8}) are also shown. 
We find that for $\mu=0.85$, the amplitudes exhibit decaying oscillations 
and approach their equilibrium value of 1, in accordance with the LV model.  
Furthermore, for $\mu=0.98$, the amplitudes exhibit large stable oscillations 
that tend to limit cycle behavior in very good agreement with the LV model's predictions.   
In addition, as seen in Fig. \ref{fig6},  
for both values of $\mu$ the soliton patterns remain intact throughout the propagation. 
Similar results are obtained for other values of $\mu$ and the $\eta_{j}(0)$.

\section{Conclusions}
\label{conclusions}
We demonstrated that the amplitudes of optical solitons 
in multisequence nonlinear waveguide coupler systems with weak linear and cubic gain-loss 
exhibit large stable oscillations along ultra-long distances. 
The stable oscillations were caused by supercritical Hopf bifurcations of the equilibrium states 
of the LV models for dynamics of soliton amplitudes. The predictions of the LV models were confirmed 
by numerical simulations with the coupled cubic NLS propagation models with $2\le N\le 4$ soliton sequences. 
Our results provide the first demonstration of intermediate 
nonlinear amplitude dynamics in multisequence soliton systems, 
described by the cubic NLS equation.
Moreover, since two of the LV models that we studied exhibit chaotic dynamics, 
our results are also an important step towards the first realization of spatio-temporal chaos 
with multiple periodic sequences of colliding NLS solitons.



\begin{thebibliography}{}
\bibitem{Zakharov84} S. Novikov, S.V. Manakov, L.P. Pitaevskii, and V.E. Zakharov, 
 {\it Theory of Solitons: The Inverse Scattering Method}  (Plenum, New York, 1984). 
 
\bibitem{Newell85} A.C. Newell, {\it Solitons in Mathematics and Physics} 
(SIAM, Philadelphia, 1985).

\bibitem{Dalfovo99} F. Dalfovo, S. Giorgini, L.P. Pitaevskii, 
and S. Stringari, Rev. Mod. Phys. {\bf 71}, 463 (1999).  

\bibitem{BEC2008} R. Carretero-Gonz\'alez, D.J. Frantzeskakis, 
and P.G. Kevrekidis, Nonlinearity {\bf 21}, R139 (2008).  

\bibitem{Agrawal2001} G.P. Agrawal, {\it Nonlinear Fiber Optics} 
(Academic, San Diego, CA, 2001).

\bibitem{Mollenauer2006} L.F. Mollenauer and J.P. Gordon,  
{\it Solitons in Optical Fibers: Fundamentals and Applications} 
(Academic, San Diego, CA, 2006).  

\bibitem{MM98} L.F. Mollenauer and P.V. Mamyshev,
IEEE J. Quantum Electron. {\bf 34}, 2089 (1998). 

\bibitem{CPN2016} D. Chakraborty, A. Peleg, and Q.M. Nguyen,
Opt. Commun. {\bf 371}, 252 (2016).  

\bibitem{PNT2016} A. Peleg, Q.M. Nguyen, and T.P. Tran, 
Opt. Commun. {\bf 380}, 41  (2016).       

\bibitem{NP2010} Q.M. Nguyen and A. Peleg, Opt. Commun. {\bf 283}, 
3500  (2010).   

\bibitem{PNC2010} A. Peleg, Q.M. Nguyen, and Y. Chung, 
Phys. Rev. A {\bf 82}, 053830 (2010).

\bibitem{PC2012} A. Peleg and Y. Chung, Phys. Rev. A {\bf 85}, 063828 (2012).

\bibitem{CPJ2013} D. Chakraborty, A. Peleg, and J.-H. Jung, 
Phys. Rev. A {\bf 88}, 023845 (2013).

\bibitem{NPT2015} Q.M. Nguyen, A. Peleg, and T.P. Tran, 
Phys. Rev. A {\bf 91}, 013839 (2015).   

\bibitem{Holmes83} J. Guckenheimer and P.J. Holmes, {\it Nonlinear Oscillations, Dynamical Systems, 
and Bifurcations of Vector Fields} (Springer,  New York, 1983). 

\bibitem{Lakshmanan2002} M. Lakshmanan and S. Rajasekar, {\it Nonlinear Dynamics} (Springer, Berlin, 2002). 

\bibitem{Field85} R.J. Field and M. Burger, eds., 
{\it Oscillations and Traveling Waves in Chemical Systems} (Wiley, New York, 1985). 

\bibitem{Murray89} J.D. Murray, {\it Mathematical Biology} (Springer, New York, 1989). 

\bibitem{Wyman89} E. Di Cera, P.E. Phillipson, and J. Wyman, 
Proc. Natl. Acad. Sci. USA {\bf 86}, 142 (1989). 

\bibitem{Odell1980} M.G. Odell, in {\it Mathematical models in molecular and 
cellular biology}, edited by L.A. Segel (Cambridge University Press, 
Cambridge, England, 1980), Appendix A.3.   

\bibitem{Arneodo80} A. Arneodo, P. Coullet, and C. Tresser, Phys. Lett. A {\bf 79}, 259 (1980).

\bibitem{Arneodo82} A. Arneodo, P. Coullet, J. Peyraud, and C. Tresser, 
J. Math. Biology {\bf 14}, 153 (1982). 

\bibitem{Namba2005} K. Tanabe and T. Namba, Ecology {\bf 86}, 3411 (2005). 

\bibitem{Vano2006} J.A. Vano, J.C. Wildenberg, M.B. Anderson, J.K. Noel, and J.C. Sprott, 
Nonlinearity {\bf 19}, 2391 (2006).  

\bibitem{Previte2013} J.P. Previte and K.A. Hoffman, SIAM Rev. {\bf 55}, 523 (2013).  

\bibitem{Lotka25} A.J. Lotka, {\it Elements of Physical Biology} 
(Williams and Wilkins, Baltimore, 1925). 

\bibitem{Volterra28} V. Volterra, J. Cons. Int. Explor. Mer {\bf 3}, 1 (1928). 

\bibitem{Wyman88} E. Di Cera, P.E. Phillipson, and J. Wyman, 
Proc. Natl. Acad. Sci. {\bf 85}, 5923 (1988). 

\bibitem{Li2008} Y. Li, H. Qian, and Y. Yi, J. Chem. Phys. {\bf 129}, 154505 (2008).

\bibitem{Agrawal2007a} Q. Lin, O.J. Painter, and G.P. Agrawal, 
Opt. Express {\bf 15}, 16604 (2007).

\bibitem{dimensions} The dimensionless distance $z$ in Eq. (\ref{HB1})  is 
$z=X/(2L_{D})$, where $X$ is the dimensional distance, 
$L_{D}=\tau_{0}^{2}/|\tilde\beta_{2}|$ is dispersion length,
$\tau_{0}$ is soliton width, and $\tilde\beta_{2}$ is the second-order
dispersion coefficient. The dimensionless time is
$t=\tau/\tau_{0}$, where $\tau$ is time. 
$\psi_{j}=E_{j}/\sqrt{P_{0}}$, where $E_{j}$ is the 
electric field of the $j$th sequence and $P_{0}$ is peak power.   
The dimensionless second-order dispersion coefficient is 
$d=-1=\tilde\beta_{2}/(\gamma P_{0}\tau_{0}^{2})$, 
where $\gamma$ is the Kerr nonlinearity coefficient.
The coefficients $\epsilon_{1}$ and $\epsilon_{3jk}$ are related 
to the dimensional linear and cubic gain-loss coefficients 
$\rho_{1}$ and $\rho_{3jk}$ by 
$\epsilon_{1}=2\tau_{0}^{2}\rho_{1}/|\tilde\beta_{2}|$  
and $\epsilon_{3jk}=2\rho_{3jk}/\gamma$. 
The solitons spectral width is $\nu_{0}=1/(\pi^{2}\tau_{0})$ and the 
intersequence frequency difference is $\Delta\nu=(\pi\Delta\beta\nu_{0})/2$. 

\bibitem{PNT2017} A. Peleg, Q.M. Nguyen, and T.T. Huynh, Eur. Phys. J. D {\bf 71}, 30 (2017).      

\bibitem{Becker99} P.C. Becker, N.A. Olsson, and J.R. Simpson, 
{\it Erbium-Doped Fiber Amplifiers: Fundamentals and Technology} 
(Academic, San Diego, CA, 1999), chapter 8.

\bibitem{mu} Note that the bifurcation parameter $\mu$ might appear 
in the expressions for the $g_{j}(z)$ or / and in the expressions for the 
cubic gain-loss coefficients $\epsilon_{3jk}$.   

\bibitem{Yang2010} J. Yang, {\it Nonlinear Waves in Integrable and Nonintegrable Systems} 
(SIAM, Philadelphia, 2010). 

\bibitem{Coste79} J. Coste, J. Peyraud, and P. Coullet, SIAM J. Appl. Math. {\bf 36}, 516 (1979). 


\end{thebibliography}
\end{document}